\documentclass[%
 reprint,
 amsmath,amssymb,
 aps,
]{revtex4-2}
\usepackage{graphicx}
\usepackage{dcolumn}
\usepackage{bm}
\usepackage{braket}
\usepackage{xcolor}
\usepackage{hyperref}
\usepackage{graphicx}
\usepackage{dcolumn}
\usepackage{bm}
\usepackage[inline]{enumitem}
\setlist{nosep}
\usepackage{amsmath,amssymb,amsfonts}

\newif\ifoutline
\outlinefalse  

\begin{document}
\preprint{APS/123-QED}
\title{Magic state cultivation on a superconducting quantum processor}
\date{\today}
\begin{abstract}
Fault-tolerant quantum computing requires a universal gate set, but the necessary non-Clifford gates represent a significant resource cost for most quantum error correction architectures. Magic state cultivation offers an efficient alternative to resource-intensive distillation protocols; however, testing the proposal’s assumptions represents a challenging departure from quantum memory experiments. We present an experimental study of magic state cultivation on a superconducting quantum processor. We implement cultivation, including code-switching into a surface code, and develop a fault-tolerant measurement protocol to bound the magic state fidelity. Cultivation reduces the error by a factor of 40, with a state fidelity of 0.9999(1) (retaining 8\% of attempts). Our results experimentally establish magic state cultivation as a viable solution to one of quantum computing's most significant challenges.
\end{abstract}
\maketitle
\onecolumngrid
\begin{flushleft}
{\small
\renewcommand{\author}[2]{#1$^\textrm{\scriptsize #2}$}
\renewcommand{\affiliation}[2]{$^\textrm{\scriptsize #1}$ #2 \\}

\newcommand{\xGoogle}{\affiliation{1}{Google Research}}
\newcommand{\xcorrespondemma}{\affiliation{*}{Corresponding author: emmarosenfeld@google.com}}
\newcommand{\xcorrespondksatz}{\affiliation{**}{Corresponding author: ksatz@google.com}}
\newcommand{\xNASA}
{\affiliation{2}{QuAIL, NASA Ames Research Center, Mountain View, CA, USA}}
\newcommand{\xUMass}{\affiliation{3}{Department of Electrical and Computer Engineering, University of Massachusetts, Amherst, MA}}
\newcommand{\xUConnStorrs}{\affiliation{4}{Department of Physics, University of Connecticut, Storrs, CT}}
\newcommand{\xAuburnECE}{\affiliation{5}{Department of Electrical and Computer Engineering, Auburn University, Auburn, AL}}
\newcommand{\xUCSB}{\affiliation{6}{Department of Physics, University of California, Santa Barbara, CA}}

\newcommand{\Google}{1}
\newcommand{\NASA}{2}
\newcommand{\UMass}{3}
\newcommand{\UConnStorrs}{4}
\newcommand{\AuburnECE}{5}
\newcommand{\UCSB}{6}
\newcommand{\correspondemma}{*}
\newcommand{\correspondksatz}{**}

\author{E. Rosenfeld}{\Google,\! \correspondemma},
\author{C. Gidney}{\Google},
\author{G. Roberts}{\Google},
\author{A. Morvan}{\Google},
\author{N. Lacroix}{\Google},
\author{D. Kafri}{\Google},
\author{J. Marshall}{\NASA},
\author{M. Li}{\Google},
\author{V. Sivak}{\Google},
\author{D. Abanin}{\Google},
% ...
\bigskip

\xGoogle
\xNASA
\xUMass
\xUConnStorrs
\xAuburnECE
\xUCSB
\xcorrespondemma
\xcorrespondksatz
}
\end{flushleft}

\clearpage
\twocolumngrid
\section{Introduction}

Non-Clifford gates are required for universal, fault-tolerant quantum computing \cite{gidney_magic_2024, bravyi_universal_2005, kubica_universal_2015, postler_demonstration_2022, dasu_breaking_2025, gupta_encoding_2024, bluvstein_architectural_2025} and indispensable for beyond-classical computation \cite{abanin_observation_2025, morvan_phase_2024, ebadi_quantum_2022}. The 2D surface \cite{bravyi_universal_2005, acharya_quantum_2025, fowler_surface_2012, eickbusch_demonstration_2025, debroy_luci_2025} and color codes \cite{bombin_gauge_2015, lacroix_scaling_2025, lee_low-overhead_2025, gidney_new_2023, kubica_universal_2015} are favorable for their planar geometry and high thresholds, but non-Clifford gates are not native. Generally, the proposed approach to implementing non-Clifford gates is to create magic states with ultra-high fidelity, then consume these states in gate teleportation \cite{rodriguez_experimental_2025, dasu_breaking_2025, daguerre_experimental_2025, gupta_encoding_2024, ye_logical_2023}. For factoring an RSA2048 key, magic state infidelities of $\sim 10^{-10}$ are required \cite{gidney_how_2025}, while infidelities between $\sim 10^{-6}$ and $ 10^{-8}$ are needed for some quantum simulations and optimization \cite{low_fast_2025, babbush_grand_2025, khattar_verifiable_2025, kan_resource-optimized_2025}. To create these ultra-high fidelity magic states, costly magic state distillation would be necessary, setting a substantial resource bottleneck for running quantum algorithms \cite{babbush_grand_2025, dalzell_quantum_2025}. 
During computation, magic states are consumed as-needed, so multiple factories can provide magic states asynchronously, allowing for error $\textit{detection}$ techniques for their generation.

Magic state cultivation \cite{gidney_magic_2024} is a recent proposal which promises to reduce this overhead, using fault-tolerant measurements of the logical state combined with post selection on the desired result. In magic state cultivation, a magic state is created in a single logical qubit with much higher fidelity than state injection protocols, which can then reduce the required rounds of magic state distillation or even eliminate it altogether \cite{chen_efficient_2025, hirano_efficient_2025, heusen_efficient_2025, vaknin_efficient_2025, gidney_magic_2024}. Its discovery has already substantially reduced estimates of the resource cost for running quantum applications \cite{gidney_magic_2024, gidney_how_2025, lee_low-overhead_2025}. 

\begin{figure*}[htbp]\label{fig1}
    \centering
    \includegraphics[width=0.67\textwidth]{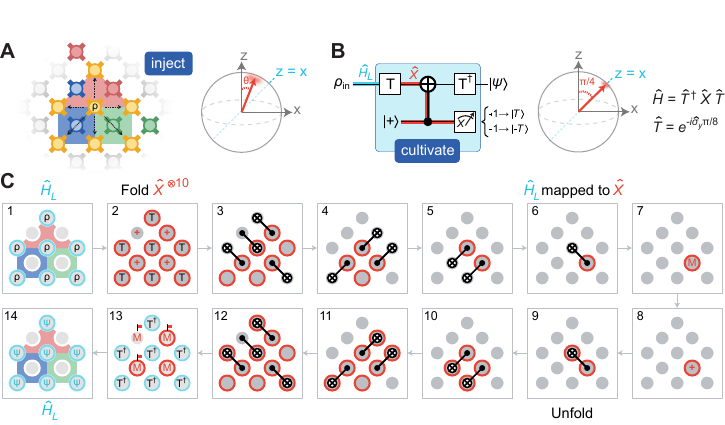}
    \caption{\textbf{Fundamentals of magic state cultivation.}
    \textbf{A}, We begin cultivation in a distance-3 color code (data qubits in yellow, helper qubits in red, blue, and green) \cite{lacroix_scaling_2025}. The color code patch sits a larger grid of transmon qubits \cite{koch_charge-insensitive_2007} of a Willow processor \cite{acharya_quantum_2025}. A state injection protocol prepares a noisy logical state $\rho$, including the color code stabilizers. 
    \textbf{B}, High-level circuit for cultivation, where we projectively measure the operator $\hat{H}_L$ on an input state $\rho_\textrm{in}$~\cite{chamberland_very_2020, gidney_magic_2024}. Quantum gates map the operator $\hat{H}_L$ into an $X$-basis measurement on the helper qubit, where post selecting for a $+1$ measurement results in a higher fidelity magic state $|\psi\rangle = |T\rangle$, projected along the desired axis.
    \textbf{C}, Detailed circuit for cultivation. Colored rings around qubits represent quantum operators (cyan: $\hat{H}$, red: $\hat{X}$) that are transformed by the quantum gates. The logical observable ($\hat{H}_L$ in step 1) is folded down to a single measurement (step 7), while flag detectors check for errors (flags, step 13). 
    The circuit reverses (steps 8-13) to recover the color code stabilizers.
    }
    \label{fig:fig1}
\end{figure*}

In this work, we present an experimental study of magic state cultivation, first implementing the protocol, then developing and executing a novel, fault-tolerant measurement scheme to characterize the refined magic state. We observe roughly a 40-fold improvement in state infidelity, relative to state-injection, corresponding to an upper bound of $1(1)\times 10^{-4}$ error, after discarding instances with detected errors in logical state preparation and measurement. About 8\% of data is retained in this case, which includes two extra rounds of cultivation and three quantum error correction (QEC) cycles. This state-of-the-art fidelity is similar to recent demonstrations in trapped ions \cite{dasu_breaking_2025, daguerre_experimental_2025}, while retaining the faster speeds afforded by superconducting circuits. Finally, to adapt our cultivated magic states into a resource for our quantum computer, we graft a magic state into a distance-5 surface code. Code-switching the state from a color code to a surface code tests the cultivation protocol beyond the practical computational limits of previous theory work \cite{gidney_magic_2024}, which simulates Clifford circuits at code distances larger than three. More importantly, grafting also begins the experimental treatment of transferring high-fidelity magic states into larger codes \cite{pogorelov_experimental_2024, butt_fault-tolerant_2024} and establishes the viability of using cultivated magic states in leading QEC architectures.

In short, magic state cultivation comprises a fault-tolerant measurement of the logical qubit along the $X = Z$, $Y = 0$ coordinates of the Bloch sphere, and post selection on the desired value. First, a noisy magic state is initialized using an injection procotol without fault-tolerance (Fig.~\ref{fig:fig1}(a), see [SI] for details), which simultaneously prepares the code stabilizers. Then, the state cultivation step projects the qubit onto the magic state axis, using phase kickback with helper qubits (Fig. 1(b), see [SI] for details). 

The phase kickback measurement constitutes a measurement of the logical Hadamard operator $\hat{H}_L$, to refine its eigenstates (i.e., a magic state). Specifically, the $\hat{H}_L$ gate is decomposed transversally onto each physical qubit as $\hat{H} = \hat{T}^\dagger \hat{X} \hat{T}$, where $\hat{T} \equiv e^{-i \pi \hat{\sigma}_y / 8}$ and $\hat{T}^\dagger$ are applied non-conditionally, leaving the conditional $\hat{X}_L$ measurement as the main task. To support this, the protocol starts in a low-distance color code (Fig.~\ref{fig:fig1}(a)), in which all the logical single-qubit gates are transversal: they can be performed by acting the operation on all physical qubits independently, limiting the spread of errors and making the $\hat{H}_L$ measurement more efficient. The logical observable component is ``folded" down to a single physical qubit for measurement (Fig.~\ref{fig:fig1}(c), purple lines). 

Subsequently, the entire process is time-reversed, which brings three advantages: the color-code stabilizers are returned back to their values at the start of the circuit, maintaining compatibility with any previous or subsequent QEC cycles, the logical observable is measured again, and the helper qubits become additional error detectors (Fig.~\ref{fig:fig1}(c), orange lines). If any detection events are registered, including a measurement of the wrong magic state, the attempt is aborted and restarted. If all checks pass, the refined magic state is then transferred into a larger code which can maintain the high state fidelity with a low logical error idling rate. In particular, the state can be ``grafted" into a surface code, favorable because of its high threshold, matching-based decoders, and well-studied dropout schemes \cite{bravyi_universal_2005, acharya_quantum_2025, fowler_surface_2012, eickbusch_demonstration_2025, debroy_luci_2025}.

Here, we study the state injection, the cultivation (i.e. phase-kickback) step, and grafting to a surface code, using a Willow processor with qubits arranged in a square lattice (Fig.~\ref{fig:fig1}(b), see [SI] for specifications). We implement the superdense color code cycle, adapted to experimental constraints following Ref.~\cite{lacroix_scaling_2025}. Throughout this work, we fine-tune our control calibration using reinforcement learning \cite{sivak_reinforcement_2025}, optimizing our control parameters by minimizing the number of error detection events in the QEC circuit. We inject the logical qubit into the $X/Z$ plane and cultivate the state $\ket{T} \equiv e^{-i \pi \hat{\sigma}_y / 8} \ket{0}$ specifically (differing from the standard $e^{-i \pi \hat{\sigma}_z/8} \ket{+}$ state by a transversal $\hat{H}_{ZY}$). This allows for tomography measurements in the $X$ and $Z$ basis, which correspond to the stabilizers of the color code, reducing the likelihood of state mis-characterization from coherent noise (see [SI] for details). During logical state tomography, the same calibration parameters are used for both axis measurements.

\section{Cultivating a magic state}

\begin{figure*}[htbp]
\centering\includegraphics[width=0.67\textwidth]{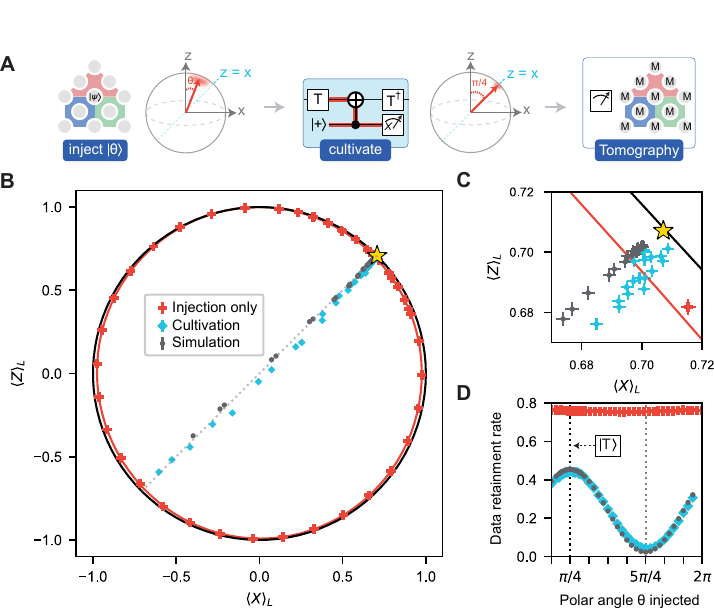}
    \caption{\textbf{Cultivating a magic state.}
    \textbf{A}, Circuit for (b, c, d). A state $\ket{\theta} \equiv e^{-i \theta \hat{\sigma}_y  /2} \ket{0}$ is first injected into the $X/Z$ plane (left Bloch sphere, red arrow). Then the cultivation circuit projects the state to the $X=Z$ line (right Bloch sphere, red arrow), and reduces the state error (red clouds).
    \textbf{B}, Resulting tomography measurements. Red crosses, without state cultivation. Cyan diamonds: with cultivation, we observe that the state is collapsed to the magic state axis (grey dashed line). Grey circles: state vector simulations, assuming a depolarizing noise model following SI1000 \cite{gidney_fault-tolerant_2021, marshall_incoherent_2025} with noise strength $p = 3.0 \times 10^{-3}$.
    \textbf{C}, Same data as in B, focused on the $\ket{T}$ state region. The cultivated state reaches closer to the desired $\ket{T}$ state (gold star) than the injected state (red curve, fit to an ellipse), which we attribute to state refinement from the cultivation step. Error bars are one standard deviation, assuming a Gaussian approximation of the binomial distribution.
    \textbf{D}, Data retainment rates, averaged over the $X$ and $Z$ tomography measurements. Cultivation adds a filter for the $\ket{T}$ state, so the retainment rates of the cultivation data and simulation show a sinusoidal dependence on $\theta$, maximized at $\theta = \pi/4$.}
    \label{fig:fig2}
\end{figure*}

We begin by comparing the logical state components before and after the cultivation step, characterized with logical state tomography (Fig.~\ref{fig:fig2}(a)). For state injection, we design a circuit which efficiently expands an arbitrary state of a single physical qubit into the logical qubit in a distance-3 color code, while also preparing all six color code stabilizers [SI]. 
We sweep over the full range of polar $\theta$ values to understand the dependence of each circuit's behavior on an arbitrary logical state $\ket{\theta} \equiv e^{-i \theta \hat{\sigma}_y/2} \ket{0}$ (in practice, to prepare $\ket{T}$, we would always set $\theta = \pi/4$). As expected, the logical state traverses the logical Bloch sphere as we vary $\theta$ (Fig.~\ref{fig:fig2}(b), red crosses). 

Motivated by the broader context of implementing magic state factories described above, we use post selection to only retain the shots of the data in which no events were observed in all of the circuit detectors. For the injection case (Fig.~\ref{fig:fig2}(b)), we retain 76(3)\% of the data (Fig.~\ref{fig:fig2}(d), red crosses), largely independent of $\theta$ because the detectors are independent of the logical qubit state. 

Next, we include the cultivation step to our circuit (Fig.~\ref{fig:fig2}(a)). We observe that the logical components follow the logical $X = Z$ line, even for $\theta \neq \pi/4$, because our $\hat{H}_L$ eigenvalue measurement for cultivation is projective. Moreover, the state components we measure are closer to $\ket{T}$ with cultivation (Fig.~\ref{fig:fig2}(c), cyan diamonds), than without it (Fig.~\ref{fig:fig2}(c), red line, elliptical fit to the injection data). We attribute this improvement to the refinement of the logical state performed by the cultivation step. 

We again post select the data for no detection events, including the phase-kickback measurement, which filters the injected logical state for the desired $\ket{T}$ state.
As expected, the data retainment rate is maximized at $\theta = \pi/4$, when encoding the state that cultivation screens for. Conversely, at the orthogonal state $\theta = 5\pi/4$, very little data is retained as cultivation removes most of it, the remainder of which we attribute to gate errors. Both the data retainment rate and state tomography curves are in approximate agreement with state vector simulations including a simple depolarizing SI1000 noise model \cite{gidney_fault-tolerant_2021} with a noise strength $p = 3.0 \times 10^{-3}$ (see [SI] for details).

We do not report a state fidelity at this stage. Foremost, a large number of measurements ($\sim 10^8$ shots) would be required to resolve errors of $\sim 10^{-4}$ within the projection noise.  Also, post selected transversal tomography can overestimate the logical state purity in the presence of coherent noise (see SI). For example, there exist pure states which appear with a Bloch vector norm greater than one, even with perfect measurement and unlimited statistics [SI]. While Pauli twirling of the noise may render the estimated state physical, in practice, twirling must be approximated. 

By definition, the cultivation circuit consists of a fault-tolerant measurement along the magic state axis. To precisely measure an upper bound on the $\ket{T}$ state error in a single, self-consistent measurement, we use cultivation itself as a  measurement device in a ``kickback tomography" (KT) approach. The circuit is shown in Fig. 3(a). First, state injection, a QEC cycle, and a first cultivation round prepares a magic state for characterization. Then, the subsequent rounds of cultivation and QEC cycles enact a fault-tolerant measurement along the $\ket{T}$ axis component to upper-bound the prepared state infidelity. 

\begin{figure*}[htbp]
    \centering
    \includegraphics[width=0.67\textwidth]{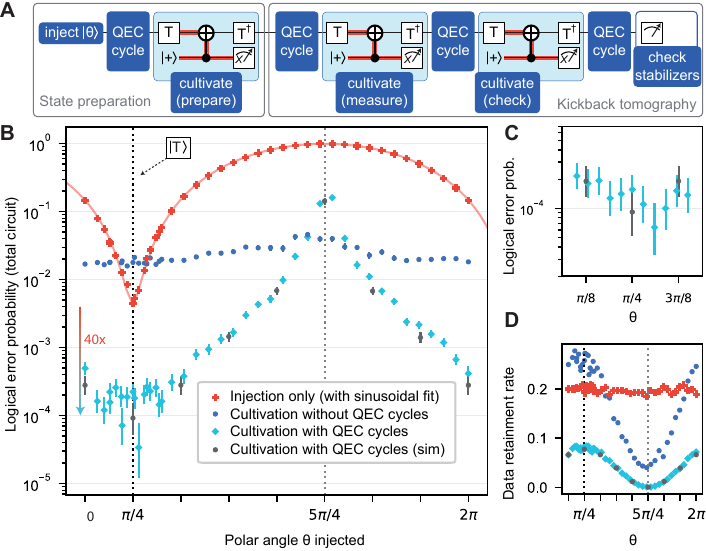}
    \caption{\textbf{Kickback tomography (KT).} 
    \textbf{A}, We prepare a $\ket{T}$ state (green box), then implement KT (brown box): subsequent rounds of cultivation self-consistently measure the $\ket{T}$ state in a single circuit that mitigates projection noise. The color code stabilizers are measured at the end for additional error detection (here we arbitrarily choose the $\hat{X}$ stabilizers).
    \textbf{B}, Logical error of the entire circuit as a function of the injected angle $\theta$. Red crosses: state preparation with injection only. Cyan diamonds: adding cultivation to the state preparation step, the logical error at $\ket{T}$ improves by about a factor of 40. Grey circles: state vector simulation including an SI1000 depolarizing noise model \cite{gidney_fault-tolerant_2021} with noise strength $p = 2.3 \times 10^{-3}$. Blue circles: removing the QEC cycles from the circuit, the logical error worsens drastically, showing the significance of the underlying color code.
    \textbf{C}, Additional data with more statistics, focused on the $\theta = \pi/4$ region of practical interest.
    \textbf{D}, Postselection rate as function of the injected angle.}
    \label{fig:fig3}
\end{figure*}
To maintain a fault distance of three, two cultivation circuits and idling circuits are needed for tomography. Intuitively, the final cultivation round acts as a ``check" of the previous one, while the idling cycles provide opportunities for measurements of the color code stabilizers, maintaining fault tolerance. 

The benefits of KT are twofold. First, since the measurements are along the desired state axis, it enables fidelity measurements in a single circuit, avoiding inconsistency issues discussed above in the presence of coherent noise. 
Second, KT reduces the projection noise compared to standard tomography: measuring along the prepared state also results in measurement averages close to zero, reducing the variance in the binomial distribution. Using KT, we are able to resolve an upper bound of the $\ket{T}$ state error of $\epsilon \sim 10^{-4}$ within an uncertainty of $\sigma \sim \epsilon$ using $\mathcal{O}(10^4)$ measurements, as opposed to $\mathcal{O}(10^8)$ as required for standard tomography. This difference enables rapid infidelity estimates in practice. 

We start by validating KT as a metrology tool, by studying the logical error of the circuit when injecting an arbitrary state on the $X/Z$ plane, removing the cultivation round, but still measuring the $\ket{T}$ component using KT. We again sweep the injected polar angle $\theta$ and measure the logical error of the total circuit and the postselection rate (Fig.~\ref{fig:fig3}(b, d), red crosses), again filtering the data for no detection events in the entire circuit. As expected, the logical error probability is sinusoidal with $\theta$, minimized when at $\theta=\pi/4$ (corresponding to $\ket{T}$), and maximized at the orthogonal state with $\theta = 5\pi/4$. Without cultivation, any error on the physical qubit used for state injection will result in a logical error. Subsequently, the logical error probability when injecting $\theta = \pi/4$ is $4.4(2) \times 10^{-3}$, similar to our physical qubit gate errors [SI]. The postselection rate is constant with $\theta$, since the detectors in this circuit do not depend on the logical qubit state.

Adding cultivation to the state preparation circuit, we observe an improvement on the logical error probability by about a factor of 40. Close to the $\ket{T}$ state injection at $\theta = \pi/4$, the logical error probability of the full circuit hovers between $1 \times 10^{-4}$ and $2 \times 10^{-4}$ (Fig.~\ref{fig:fig3}(b, c), cyan diamonds). This error probability is well below the error of each of the 216 physical two-qubit gates in the circuit, as well as its 54 measurements [SI].
Filtering for any detection events in the entire circuit, the data retainment rate is maximized again at 8\% at $\theta = \pi/4$, and minimized at $\theta = 5\pi/4$ (Fig.~\ref{fig:fig3}(d)). These results are in agreement with state vector simulations of the circuit, as shown in Fig.~\ref{fig:fig3}(b, c), grey dots, which simulate the cultivation circuit with a simple SI1000 depolarizing noise model \cite{gidney_fault-tolerant_2021} with a noise strength $p = 2.3 \times 10^{-3}$. The difference in the device noise strength $p$ compared to that of Fig.~\ref{fig:fig2} is likely due to changes in optimal qubit frequencies and the distinct control calibrations, while device drift results in logical error variation on the order of $1 \times 10^{-4}$ over about an hour [SI].

The cultivation circuit does not perform a measurement of the color code stabilizers. Therefore, when we remove the QEC cycles, we remove the quantum error detection framework that upholds the fault distance of magic state cultivation. In particular, with only one opportunity for stabilizer measurements (at the very end), there are single error events that can cause a logical error. In this case, removing the QEC cycles makes the error about 100 times worse (blue dots, Fig.~\ref{fig:fig3}(b)). 
The color code is effective: \textit{adding} four QEC cycles, each with 30 two-qubit gates and 6 measurements, improves logical performance by about a factor of 100. 
\begin{figure*}[htbp]
    \centering
    \includegraphics[width=0.67\textwidth]{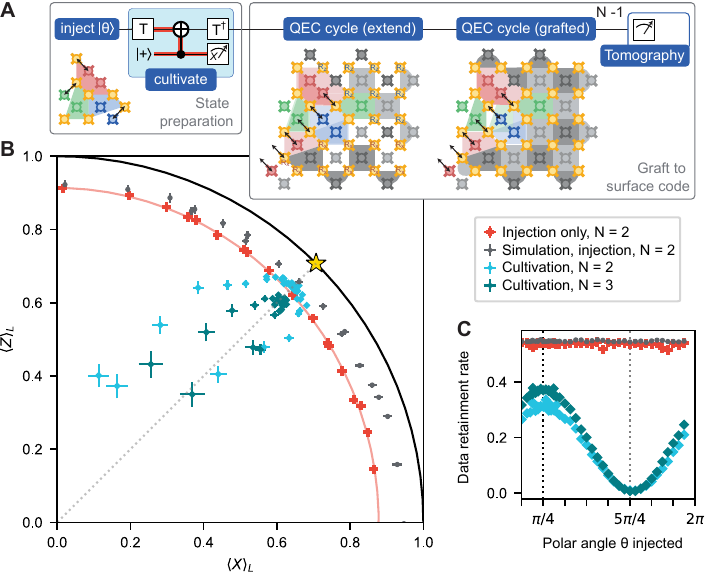}
    \caption{\textbf{Grafting from a color code to a surface code.}
    \textbf{A}, We prepare a state in our distance-3 color code (left qubit layout) and then expand to a larger code using additional qubits in an extension cycle (central qubit layout). The extension cycle also changes the pairs of data and measure qubits we exchange (black arrows), for compatibility between our $d=3$ color code geometry and the stabilizers in \cite{gidney_magic_2024}. For a total of $N$ cycles in the larger grid, we subsequently run $N-1$ cycles in the grafted code, at which point all the surface code stabilizers are prepared (right-most qubit layout). The state is then characterized with logical state tomography. 
    \textbf{B}, Logical tomography results. Red crosses: the injection-only state preparation results traverse a circle, as expected. Grey points: state vector simulations of the injection-only circuit, over all 54 qubits. Cyan and dark cyan diamonds: adding cultivation to state preparation, the states are projected toward the $Z=X$ line (see [SI] for simulations), for $N=2$ and $N=3$.
    \textbf{C}, Data retainment rate as a function of $\theta$. See text for details on post selection.} 
    \label{fig:fig4}
\end{figure*}

\section{Grafting to a surface code}
The surface code boasts a high threshold, matchable decoders, and well-developed dropout approaches \cite{bravyi_universal_2005, acharya_quantum_2025, fowler_surface_2012, eickbusch_demonstration_2025, debroy_luci_2025}, but the color code supports cultivation through its transversal single-qubit Clifford gates \cite{gidney_magic_2024}. We demonstrate end-to-end compatibility of our magic states with the surface code, by grafting our magic states into a distance-5 encoding compatible with the surface code for subsequent lattice surgery (Fig.~\ref{fig:fig4}(b)) \cite{gidney_magic_2024, butt_fault-tolerant_2024, heusen_efficient_2025, pogorelov_experimental_2024}. In a small code, the high error per QEC cycle following cultivation engulfs any magic state fidelity improvements from cultivation \cite{gidney_magic_2024}, so the state must escape into a larger code. 

We transition from the color code into the larger grafted code in two steps (Fig.~\ref{fig:fig4}(a), see [SI] for the full circuit). First, a single cycle shifts our original color code to an alternate geometry compatible with the surface code stabilizers described in \cite{gidney_magic_2024}, while simultaneously preparing the stabilizers in the surface code region (Fig.~\ref{fig:fig4}(a), QEC cycle (extend)). The data qubits introduced in the surface code region are initialized into the basis that maximizes the number of detectors in this initial cycle, as shown in the central qubit layout. To idle in the larger grid for a total of $N$ cycles, we then run $N-1$ cycles in this grafted code. In implementation, $d-1$ cycles would be needed before switching to a matchable code, beyond the scope of this work.

For consistency with Figs.~\ref{fig:fig2} and \ref{fig:fig3}, we post select for no detection events on the state preparation detectors before grafting. The grafted code is not matchable, as it contains stabilizers similar to the color code as well as surface code (Fig.~\ref{fig:fig4}(a), qubit grids), so we decode the resulting data with Tesseract, a most likely error decoder using the $A^*$ search technique \cite{beni_tesseract_2025}. 

We demonstrate grafting by characterizing the logical qubit components in the grafted code, with and without cultivation as part of the state preparation step (Fig.~\ref{fig:fig4}(a), green box). We begin by injecting $\ket{\theta}$, running one extension cycle and one cycle in the grafted code, and finally performing tomography. Sweeping $\theta$, we observe a circle on the logical Bloch sphere in the $XZ$ plane, as expected (Fig.~\ref{fig:fig4}(b), red crosses). State vector simulations across all 54 qubits, again using SI1000 \cite{gidney_fault-tolerant_2021} with $p = 2.3 \times 10^{-3}$, are shown in Fig.~\ref{fig:fig4}(b, c), grey points (see [SI] for simulation details). The difference between the experiment and simulation may be due to population leakage, which is not included in our noise model [SI], or from the variability in physical qubit performance. 

Next, we graft a cultivated magic state into the surface code, by adding the cultivation step before grafting (Fig.~\ref{fig:fig4}(b, c), cyan diamonds). These results test magic state cultivation beyond the Pauli approximations of \cite{gidney_magic_2024} and confirm that cultivated magic states may be transferred to larger codes. The state becomes compressed along the $\ket{T}$ axis, similar to Fig.~\ref{fig:fig2} (see [SI] for simulations). In this case, the data does not extend past the origin of the Bloch sphere towards the orthogonal state. We attribute this to the different logical error mechanisms that can occur when the circuit structure is changed, and the effect is reproduced in simulations [SI]. The spread from the $Z=X$ may be due to time-correlated phase noise in the circuit, and is a topic for future study.
As expected, the distance to the surface of the Bloch sphere at the $\ket{T}$ state is about the same as the injection-only case, since both are limited by the idle error in the $d=5$ grafted code (for a full memory experiment in this code, see [SI]).
 
Adding a cycle of the grafted code (Fig.~\ref{fig:fig4}(c), dark cyan diamonds) moves the state closer to the origin of the Bloch sphere, because of the finite error per cycle. While the data retainment rate should be independent of $N$ for our post selection strategy (see [SI] for simulations), we observe a ~10\% difference in data retainment rates around $\theta = \pi/4$ for the $N = 2$ and $N = 3$ experiments. This difference could be due to flux transients in our circuit, or drift in the optimal calibration parameters over time (see [SI]). Overall, our grafting results show a proof-of-principle demonstration of non-Clifford state escape into an advantageous code.

\section{Conclusion and outlook}
Our experiment establishes magic state cultivation as a low-cost technique for generating high-fidelity magic states. Looking ahead, as the processor noise strength $p$ \cite{gidney_fault-tolerant_2021} improves, the magic state infidelity should drastically scale as $p^3$. For example, improving the noise strength by a factor of two should result in $\ket{T}$ state infidelity upper bounds of $\sim 2 \times 10^{-5}$ [SI], approaching relevance for some applications \cite{babbush_grand_2025, khattar_verifiable_2025, kan_resource-optimized_2025, low_fast_2025} while using only a distance-3 code for the cultivation step. Beyond device and logical state performance improvements, future directions include real-time implementation of $\ket{T}$ gates, grafting into larger distances to support high state fidelites, further study into decoders for the grafted code \cite{beni_tesseract_2025, bausch_learning_2024}, and implementation of a matchable code following grafting as in \cite{gidney_magic_2024}. 

Our results also highlight that experimental quantum error correction research has utility for understanding logical circuits. While standard logical state tomography confirms that cultivation entails a strong, projective measurement of $\hat{H}$, we find that to quantify our state performance rigorously, we need to develop and implement additional quantum error correction techniques. Specifically, by using the fault tolerance of the cultivation step and stabilizer checks in the superdense color code cycle, we can arrive at a precise lower bound on the magic state fidelity. Moreover, a standard approximation in QEC theoretical studies is to replace non-Clifford states with Pauli states, for computational speed. Our work demonstrates that experimental QEC research can study these assumptions, as exemplified by grafting magic states into the $d=5$ surface code, extending beyond the simulations of \cite{gidney_magic_2024}. Ultimately, these findings transform magic state cultivation from a theoretical study into an experimental reality, providing a practical route to universal logic in superconducting processors. 

\textbf{Acknowledgments and contributions.} E.R. constructed the circuits, performed the experiments, simulations, and analyses. C.G. provided mentorship on circuit construction and the cultivation process. E.R. and C.G. designed the kickback tomography protocol and the injection circuit. K.J.S. conceived the study and directed the research. A.~Bourassa, G.R., N.L., and A.M. provided feedback and experimental support throughout the project.
The Google Quantum AI team sponsored the project, designed and fabricated the device, built and maintained the cryogenic and control systems, provided software infrastructure, performed the initial bring up of the device, and provided relevant guidance.

\textbf{Data availability.} Data is available at \href{https://doi.org/10.5281/zenodo.17872003
}{https://doi.org/10.5281/zenodo.17872003}
\bibliography{bib}

\renewcommand{\thefigure}{S\arabic{figure}}

\begin{titlepage}
    \centering
    {\bfseries Supplemental material to ``Magic state cultivation on a superconducting quantum processor" \par}
    \vspace{0.5cm}
\end{titlepage}
\onecolumngrid
\tableofcontents

\section{\label{sec:device}Willow device properties}
\begin{figure*}[ht]
    \centering \includegraphics[width=\textwidth]{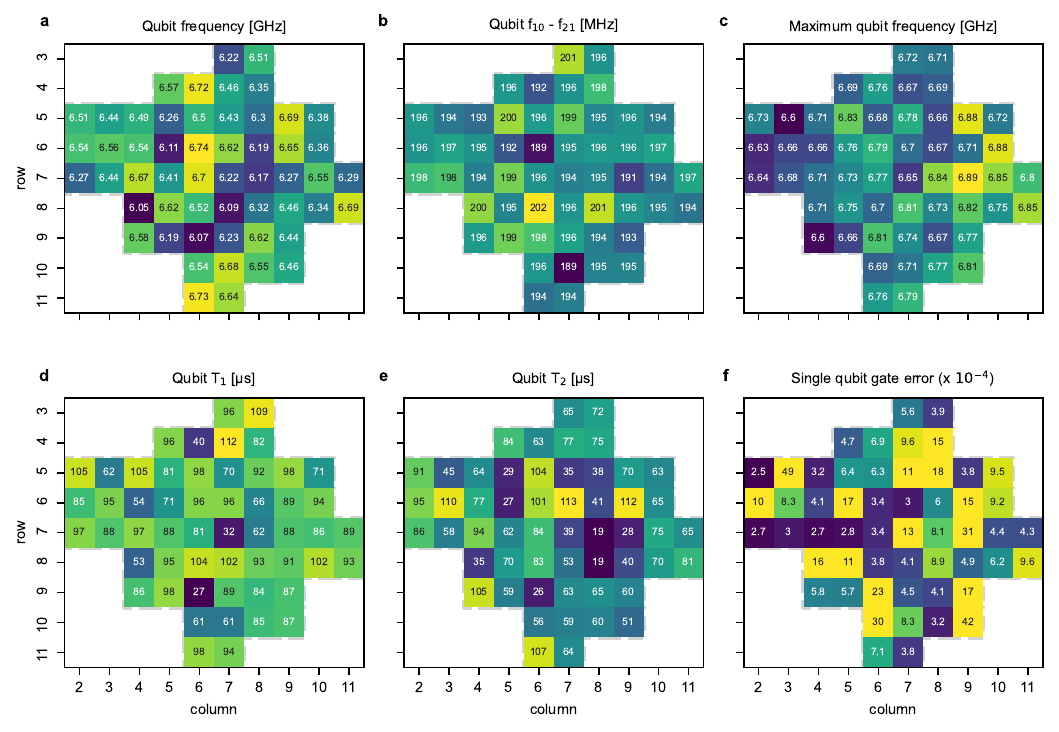}
    \caption{Single-qubit specifications. The qubits are frequency-tunable; performance numbers are specific to the indicated operating frequencies for Fig. 4 of the main text. The coherence $T_2$ is from a CPMG experiment. Single-qubit gate error is from Clifford randomized benchmarking (including both $\pi$ and $\pi/2$ rotations), and we report ``average" error. The single-qubit gates all had duration 35~ns.}
    \label{fig:qubit_specs}
\end{figure*}
\begin{figure*}[ht]
    \centering \includegraphics[width=\textwidth]{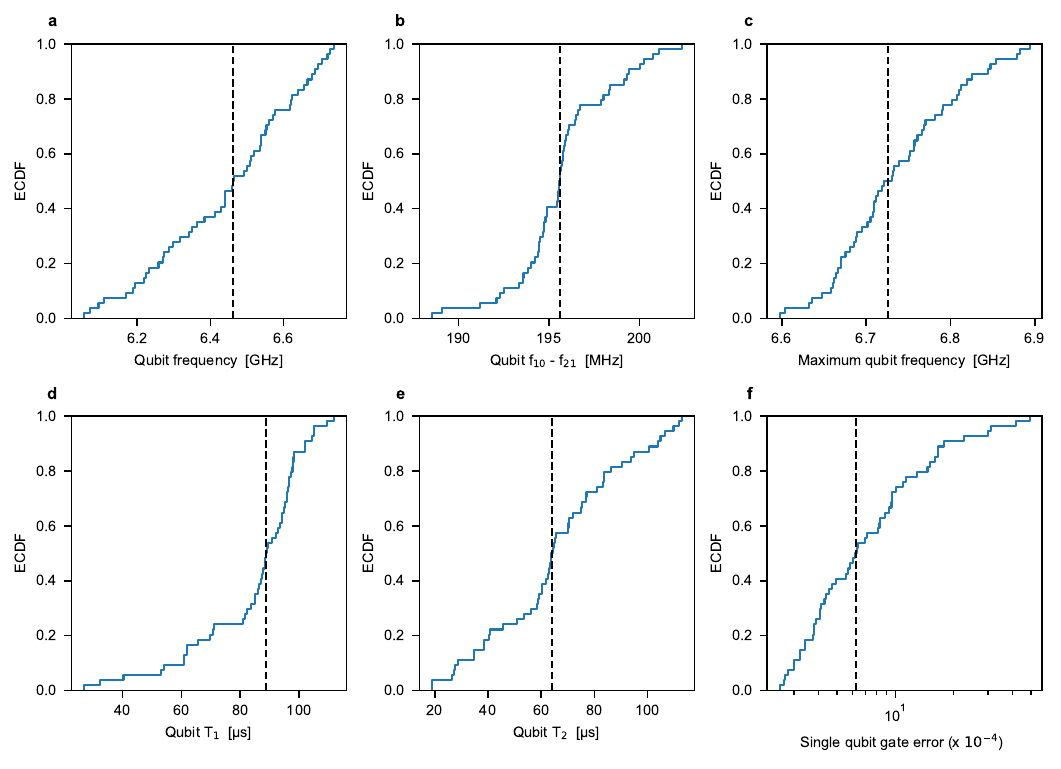}
    \caption{Empirical cumulative distribution functions (EDCF) of single-qubit specifications. Median values are shown with black dashed lines.}
    \label{fig:qubit_specs_ecdf}
\end{figure*}
\begin{figure*}[htbp]
    \centering \includegraphics[width=\textwidth]{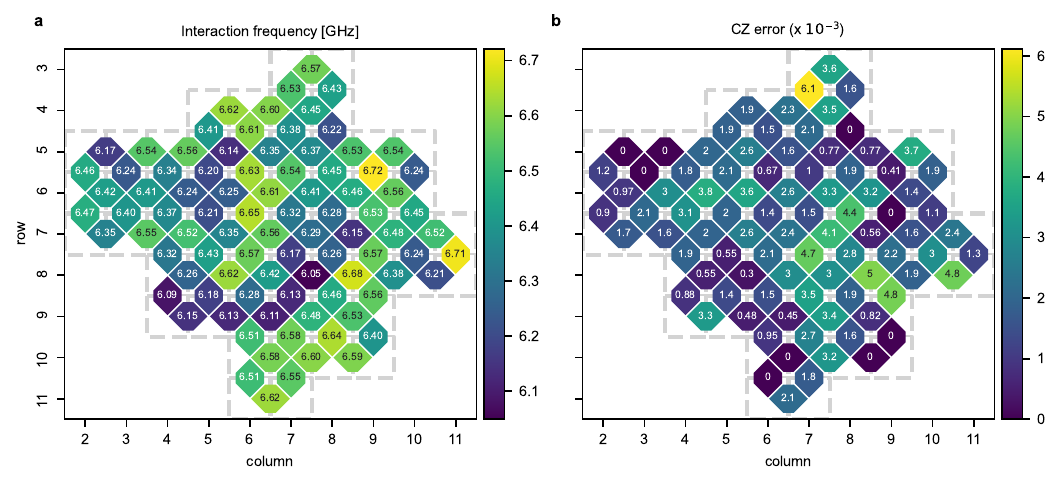}
    \caption{Two-qubit CZ gate specifications on our device, for the control parameters used in Fig. 4 of the main text. The CZ gates all had duration 35~ns.}
    \label{fig:pair_specs}
\end{figure*}
\begin{figure*}[htbp]
    \centering \includegraphics[width=\textwidth]{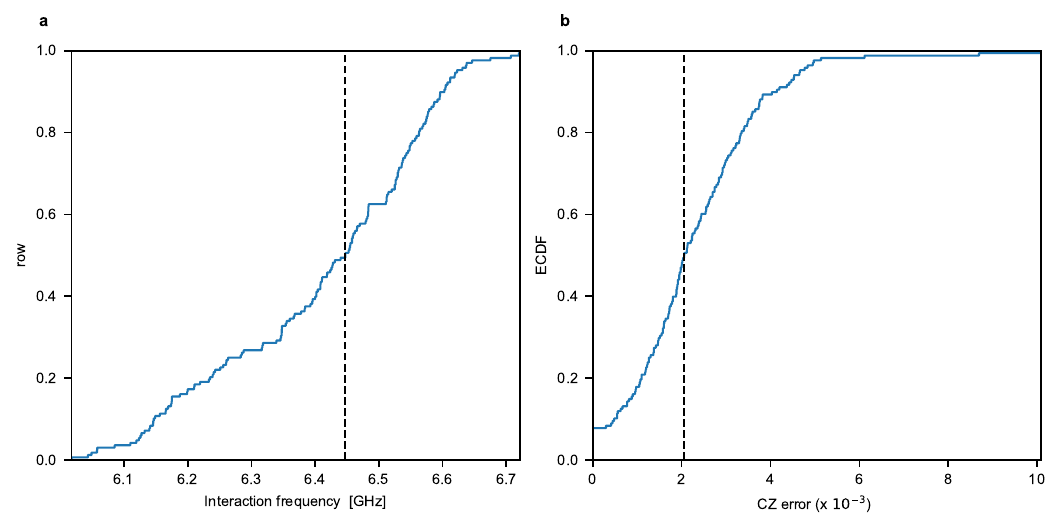}
    \caption{Empirical cumulative distribution functions (EDCF) of CZ gate specifications. Median values are shown with black dashed lines.}
    \label{fig:pair_specs_ecdf}
\end{figure*}
\begin{figure*}[htbp]
    \centering \includegraphics[width=\textwidth]{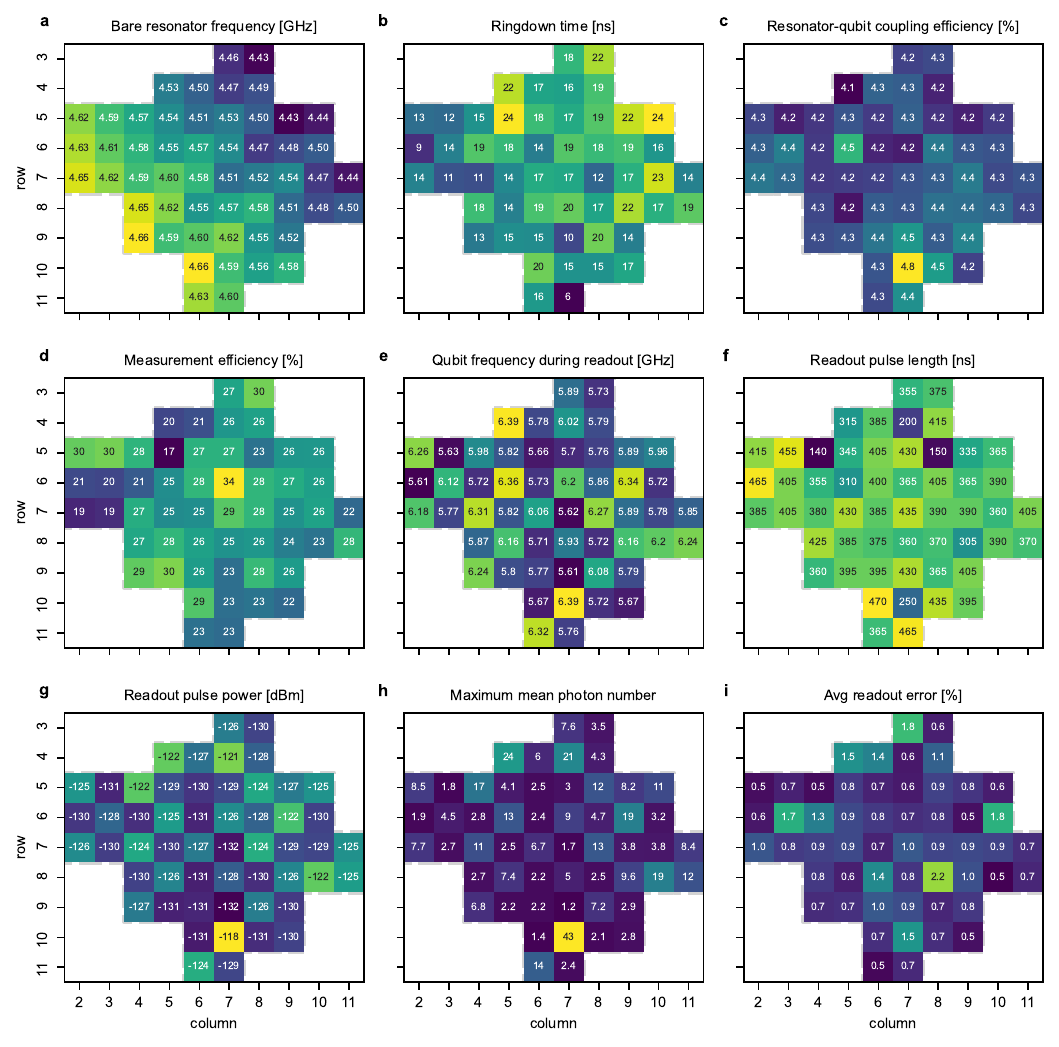}
    \caption{Readout performance specifications. Qubits and control parameters correspond to the ones used in Fig. 4 of the main text.}
    \label{fig:ro_specs}
\end{figure*}
Our work is performed on a Willow processor~\cite{acharya_quantum_2025}, which contains 105 tunable transmon qubits in a square lattice, coupled by tunable transmon couplers. The specifications for device performance are similar to Ref~\cite{acharya_quantum_2025}.  We plot the device single-qubit parameters below in Fig.~\ref{fig:qubit_specs}, and the cumulative distributions in Fig.~\ref{fig:qubit_specs_ecdf}. We include qubits that were active during the grafting experiment in Fig. 4 of the main text, the largest qubit lattice of this work. The CZ two-qubit gate interaction frequencies (the midpoint frequencies of the qubits during the gate) and CZ fidelities are shown in Fig.~\ref{fig:pair_specs}. Cumulative distributions for those are shown in Fig.~\ref{fig:pair_specs_ecdf}. 
Readout and reset parameters are shown in Fig.~\ref{fig:ro_specs}. These parameters are taken from the experiment shown in Fig.~2 of the main text. 

\section{Additional description of magic state cultivation}
\begin{figure*}[ht!]
\centering
\includegraphics[width=0.6\linewidth]{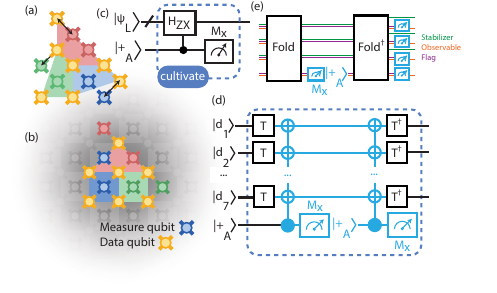}
    \caption{Details of magic state cultivation. (a) State injection in a color code. We swap data and measure qubit pairs (black arrows) to satisfy readout line constraints. (b) Patch of d=3 color code qubits on a Willow processor. (c) Phase kickback high level description, indicating the projective measurement of the logical Hadamard operator. (d) The conditional $\hat{H}_L$ gate is decomposed into its transversal physical qubit gates ($\hat{H}^{\otimes 7}$, which is in turn decomposed into $\hat{T}$ and $\hat{X}$ gates using the identity $\hat{H} = \hat{T} \hat{X} \hat{T}^\dagger$, where the $\hat{X}$ is conditional on a helper qubit initialized in $|+\rangle$. The conditional $\hat{X}$ measurement is repeated in a ``double-check" of the logical state component. (e) Graphical depiction of the logical observable and stabilizers in MSC during the cyan part of the circuit in (d). By repeating the logical state measurement in a time reverse of the first, the stabilizers (pink) return to their original value, where they may be used for subsequent QEC cycles. Additional flag detectors (purple) detect additional error mechanisms and constitute the second observable measurement, which together brings the fault distance of the circuit to three. For a detailed description of how the second observable measurement and flag checks work in the circuit, see Fig.~\ref{fig:cultivation_flags}.}
    \label{fig:cultivation_basics}
\end{figure*}

\begin{figure*}[ht!]
\centering
\includegraphics[width=0.8\linewidth]{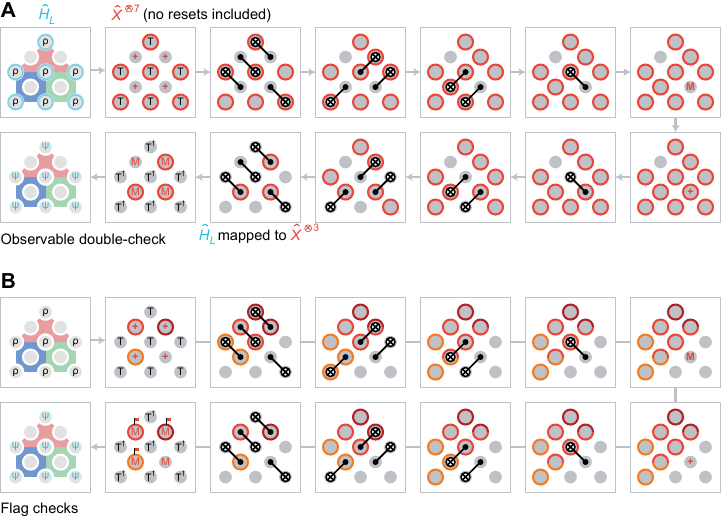}
    \caption{Flag and double-check detectors in magic state cultivation. Colored outlines on the qubits represent operators being transformed by the gates of the circuit (cyan: $\hat{H}$, red: $\hat{X}$). Measurements are in the $X$ basis. (a) The second observable measurement results at the final measurements in the circuit, just after the $\hat{H}$ operator is mapped to three $X$ components on the flag qubits. (b) Because the cultivation circuit is time-symmetric, the measurements at the end are deterministic (in the absence of errors). This results in ``flag detectors" which help to check if errors have occurred during cultivation. We follow the three flag checks through the circuit with different red and orange shading. }
    \label{fig:cultivation_flags}
\end{figure*}

Cultivation includes a fault-tolerant measurement along the magic state axis, with post selection on the desired result. The conditional, logical $\hat{H}_L$ gate as required for phase kickback occurs through a  decomposition of the transversal $\hat{H}_L$ gate into physical $\hat{T} \equiv e^{-i \hat{\sigma}_y \pi/8}$ and conditional $\hat{X}$ (CNOT) operations. First, $\hat{T}$ gates are applied to every data qubit. This causes the processor to temporarily leave the color code, as the state is no longer an eigenstate of the stabilizers following the $\hat{T}$ gates. Then, a conditional $\hat{X}$ is performed by folding the joint parity of the data qubits' $\hat{X}$ component down to a single qubit, which is measured (Fig. \ref{fig:cultivation_basics}(d), circuit up to first $M_x$). To restore the color code stabilizers, the process is time reversed, which simultaneously acts as a second measurement of the logical $\hat{X}$ component (Fig. \ref{fig:cultivation_basics}(d), second $M_x$), as well as adds flag checks to catch additional errors. The second observable and flag checks are crucial for achieving fault tolerance; to follow how they work within the circuit, see Fig.~\ref{fig:cultivation_flags}, which highlights how the relevant operators translate into measurements. Because the code stabilizers are ultimately maintained through this process (Fig. \ref{fig:cultivation_basics}(e)), errors on physical qubits can also be detected and post selected with subsequent stabilizer measurements. Finally, after post selection on the measurement results for the desired $\ket{T}$ state, the ``cultivation" step is complete, and the state becomes ``cultivated." This fault-tolerant ``cultivation” of the logical observable extends the fault distance of the entire circuit to three, meaning specific chains of three physical qubit errors are required to result in a logical error (as opposed to state injection which generally has a fault distance of one, where single errors on the injection qubit propagate out to logical errors).

\section{\label{sec:explicit_circuits}Explicit circuits}
Here, we attach crumble links for the explicit circuits used for each element reported in the main text. Detectors are included in the circuit and shown as small colored squares in the dots. For the crumble manual, see \href{https://github.com/quantumlib/Stim/blob/main/glue/crumble/README.md}{here} (use keyboard ``e" to move forward and ``q" to move back through the circuit). Circuits were constructed using \hyperlink{https://github.com/quantumlib/Stim/tree/main}{STIM}. To access the STIM circuit, click ``Show Import/Export." In the following links, we include an example stabilizer and/or observable component for illustrative purposes. 
\begin{itemize}
\item \href{https://algassert.com/crumble#circuit=Q(2,2)0;Q(3,1)1;Q(3,2)2;Q(3,3)3;Q(4,1)4;Q(4,2)5;Q(4,3)6;Q(4,4)7;Q(5,1)8;Q(5,2)9;Q(5,3)10;R_1_2_3_6_7_8_5_4_9_10_0;MARKZ(0)5;MARKZ(1)9;POLYGON(0,0,1,0.25)1_8_10_5;POLYGON(0,1,0,0.25)1_5_3_0;POLYGON(1,0,0,0.25)5_10_7_3;TICK;H[injection]_5;TICK;I[injection]_5;TICK;H_0_1_2_4_8_9_10;TICK;CZ_1_4_2_5_8_9;TICK;H_2_3_8;TICK;CZ_2_3;TICK;H_3_5_6;TICK;CZ_1_2_3_6_4_5;TICK;H_2_6;TICK;CZ_2_3_5_6;TICK;H_6;TICK;CZ_0_2_6_10;TICK;H_2_3_4_6_7;TICK;CZ_1_2_3_6;TICK;CZ_2_5_4_8_6_7;TICK;H_4_7_10;TICK;CZ_1_4_2_3_5_6_9_10;TICK;H_1_2_3_6_9_10;TICK;CZ_0_2_4_5_6_7_8_9;TICK;H_2_4_5_6_9;TICK;M_2_4_9_6;MARKZ(1)4}{State injection circuit} 
\item \href{https://algassert.com/crumble#circuit=Q(2,2)0;Q(3,1)1;Q(3,2)2;Q(3,3)3;Q(4,1)4;Q(4,2)5;Q(4,3)6;Q(4,4)7;Q(5,1)8;Q(5,2)9;Q(5,3)10;Q(6,2)11;R_2_4_6_9;MARKX(0)7_3_0_1_5_10_8;MARKX(1)10_5_1_8;MARKZ(0)2_6_9;POLYGON(0,0,1,0.25)1_8_10_5;POLYGON(0,1,0,0.25)1_5_3_0;POLYGON(1,0,0,0.25)5_10_7_3;TICK;I[T_gate]_0_1_3_5_7_8_10;TICK;H_1_2_4_5_6_7_8_9;TICK;CZ_1_2_4_5_6_7_8_9;TICK;H_0_5_9_10;TICK;CZ_0_2_5_9_6_10;TICK;H_2_3_9_11;TICK;CZ_2_5_3_6_9_11;TICK;H_5_9_11;TICK;CZ_5_6_9_11;TICK;H_6_9;TICK;M[root_measurement]_6_9;MARKZ(0)6;TICK;R_6_9;TICK;H_6_9;TICK;CZ_5_6_9_11;TICK;H_5_9_11;TICK;CZ_2_5_3_6_9_11;TICK;H_2_3_9_11;TICK;CZ_0_2_5_9_6_10;TICK;H_0_5_9_10;TICK;CZ_1_2_4_5_6_7_8_9;TICK;H_1_2_4_5_6_7_8_9;TICK;I[negative_T_gate]_0_1_3_5_7_8_10;TICK;M_2_4_6_9}{Cultivation circuit}
\item \href{https://algassert.com/crumble#circuit=Q(0,2)0;Q(0,3)1;Q(0,4)2;Q(1,2)3;Q(1,3)4;Q(1,4)5;Q(2,2)6;Q(2,3)7;Q(2,4)8;Q(2,5)9;Q(2,6)10;Q(3,1)11;Q(3,2)12;Q(3,3)13;Q(3,4)14;Q(3,5)15;Q(3,6)16;Q(4,1)17;Q(4,2)18;Q(4,3)19;Q(4,4)20;Q(4,5)21;Q(4,6)22;Q(4,7)23;Q(4,8)24;Q(5,0)25;Q(5,1)26;Q(5,2)27;Q(5,3)28;Q(5,4)29;Q(5,5)30;Q(5,6)31;Q(5,7)32;Q(5,8)33;Q(6,0)34;Q(6,1)35;Q(6,2)36;Q(6,3)37;Q(6,4)38;Q(6,5)39;Q(6,6)40;Q(6,7)41;Q(7,2)42;Q(7,3)43;Q(7,4)44;Q(7,5)45;Q(7,6)46;Q(7,7)47;Q(8,2)48;Q(8,3)49;Q(8,4)50;Q(8,5)51;Q(9,4)52;Q(9,5)53;POLYGON(0,0,1,0.25)11_18_13_6;POLYGON(0,1,0,0.25)18_28_20_13;POLYGON(1,0,0,0.25)26_28_18_11;TICK;MPP_Z11*Z18*Z13*Z6;TICK;MPP_X11*X18*X13*X6;TICK;MPP_Z18*Z28*Z20*Z13;TICK;MPP_X18*X28*X20*X13;TICK;MPP_Z26*Z28*Z18*Z11;TICK;MPP_X26*X28*X18*X11;TICK;R_44_27_42_52_29_12_14_21_41_5_9_23_36_43_45_48_50_53_38_40_47_34_3;RX_51_37_17_25_35_49_19_39_7_31_1_16_33_46_0_8_15_22_10_2_24_32_30_4;MARKX(0)13_18_26_17_25_1_7_8_15;POLYGON(0,0,1,0.25)18_28_20_13;POLYGON(0,1,0,0.25)11_18_13_6;POLYGON(1,0,0,0.25)26_28_18_11;TICK;CX_35_34_25_26_49_48_43_42_53_52_37_36_45_44_17_11_19_18_39_38_20_29_46_40_30_21_47_41_8_14_15_9_31_22_32_23_16_10_33_24_51_50_2_5_0_1;TICK;CX_34_35_26_25_48_49_42_43_52_53_36_37_44_45_11_17_38_39_29_20_40_46_21_30_41_47_14_8_9_15_22_31_23_32_10_16_24_33_50_51_5_2;TICK;CX_34_25_26_27_11_12_48_42_36_37_19_20_7_8_1_2_38_39_10_9_24_23_46_47_52_53;TICK;CX_34_35_26_25_48_49_27_28_12_13_20_21_8_9_10_16_24_33_51_53_41_47;TICK;CX_26_17_36_42_27_18_12_6_37_28_19_13_7_4_38_29_20_14_8_5;TICK;CX_11_17_36_35_12_18_49_43_19_28_50_52_38_44_20_29_8_14_51_45_39_30_21_15_40_46_22_31_41_32;TICK;CX_25_26_18_19_6_7_28_27_13_12_50_49_44_43_38_37_46_45_40_39_31_30_22_21_16_15_33_32;TICK;CX_17_26_42_43_18_27_6_12_37_28_13_19_4_7_50_51_44_45_29_30_14_15_40_41_31_32_22_23;TICK;CX_35_26_17_11_18_12_28_19_13_7_4_1_29_20_14_8_5_2;TICK;CX_26_27_11_12_37_43_19_20_7_8_1_2_50_44_39_45_21_30_9_15_40_31_22_16_23_32;TICK;CX_28_27_19_18_13_12_7_6_4_3_1_0;TICK;CX_27_28_18_19_12_13_6_7_3_4_0_1;TICK;M_43_20_8_2_45_30_15_32_47_53_28_13_4;MX_34_26_11_48_36_50_38_40_22_10_24_18_0_6;DT(7,3,0)rec[-27];DT(4,2,0)rec[-26]_rec[-31];DT(7,5,0)rec[-23];DT(7,7,0)rec[-19];DT(9,5,0)rec[-18];DT(5,1,0)rec[-17]_rec[-29];DT(1,3,0)rec[-15];DT(3,1,0)rec[-15]_rec[-16]_rec[-33];DT(5,3,0)rec[-13]_rec[-28];DT(3,3,0)rec[-12]_rec[-32];DT(4,6,0)rec[-6];DT(2,6,0)rec[-5];DT(4,8,0)rec[-4];DT(4,4,0)rec[-3]_rec[-30];DT(0,2,0)rec[-2];DT(2,2,0)rec[-1];TICK;MPP_Z25*Z35*Z37*Z27*Z19*Z17_X3*X14*X9*X5_X49*X42_Z52*Z51_X39*X46*X41*X31_X33*X23;DT(5,0,1)rec[-6];DT(0,2,1)rec[-5]_rec[-8];DT(8,2,1)rec[-4]_rec[-17];DT(9,5,1)rec[-3]_rec[-24];DT(6,6,1)rec[-2]_rec[-13];DT(4,8,1)rec[-1]_rec[-10];TICK;MPP_X12*X19*X27*X29*X21*X14_Z5*Z3_X35*X25_Z44*Z51*Z46*Z39_Z31*Z41*Z33*Z23_X16*X9;DT(4,2,2)rec[-6]_rec[-15];DT(1,3,2)rec[-5]_rec[-27]_rec[-36];DT(6,0,2)rec[-4]_rec[-26];DT(9,5,2)rec[-3]_rec[-30]_rec[-35];DT(5,7,2)rec[-2]_rec[-32];DT(2,6,2)rec[-1]_rec[-17];TICK;MPP_X25*X27*X19*X17_Z12*Z14*Z7*Z3_X49*X52*X51*X44_X35*X42*X37_Z41*Z46_Z29*Z39*Z31*Z21;DT(4,2,3)rec[-6]_rec[-21]_rec[-30]_rec[-31];DT(3,3,3)rec[-5]_rec[-34]_rec[-43];DT(8,4,3)rec[-4]_rec[-27];DT(6,2,3)rec[-3]_rec[-28]_rec[-31];DT(7,7,3)rec[-2]_rec[-37];DT(5,5,3)rec[-1]_rec[-40];TICK;MPP_Z17*Z19*Z7*Z12_X27*X37*X44*X39*X29_Z14*Z21*Z16*Z9;DT(4,1,4)rec[-3];DT(4,2,4)rec[-2]_rec[-24]_rec[-29];DT(3,5,4)rec[-1]_rec[-42];TICK;MPP_Z27*Z29*Z19*Z12_Z42*Z49*Z44*Z37_X21*X31*X23*X16;DT(3,3,5)rec[-3]_rec[-40]_rec[-41]_rec[-50];DT(7,3,5)rec[-2]_rec[-51];DT(4,6,5)rec[-1]_rec[-30];TICK;MPP_X17*X19*X7*X12;DT(2,2,6)rec[-1]_rec[-26]_rec[-28]_rec[-38]}{Grafting - adapt to new code}
\item \href{https://algassert.com/crumble#circuit=Q(0,2)0;Q(0,3)1;Q(0,4)2;Q(1,2)3;Q(1,3)4;Q(1,4)5;Q(2,2)6;Q(2,3)7;Q(2,4)8;Q(2,5)9;Q(2,6)10;Q(3,1)11;Q(3,2)12;Q(3,3)13;Q(3,4)14;Q(3,5)15;Q(3,6)16;Q(4,1)17;Q(4,2)18;Q(4,3)19;Q(4,4)20;Q(4,5)21;Q(4,6)22;Q(4,7)23;Q(4,8)24;Q(5,0)25;Q(5,1)26;Q(5,2)27;Q(5,3)28;Q(5,4)29;Q(5,5)30;Q(5,6)31;Q(5,7)32;Q(5,8)33;Q(6,0)34;Q(6,1)35;Q(6,2)36;Q(6,3)37;Q(6,4)38;Q(6,5)39;Q(6,6)40;Q(6,7)41;Q(7,2)42;Q(7,3)43;Q(7,4)44;Q(7,5)45;Q(7,6)46;Q(7,7)47;Q(8,2)48;Q(8,3)49;Q(8,4)50;Q(8,5)51;Q(9,4)52;Q(9,5)53;POLYGON(0,0,1,0.25)12_29_21_14;POLYGON(0,0,1,0.25)12_29_21_14;POLYGON(0,0,1,0.25)12_29_21_14;POLYGON(0,0,1,0.25)27_29_19_12;POLYGON(0,0,1,0.25)27_29_19_12;POLYGON(0,0,1,0.25)27_29_19_12;POLYGON(0,0,1,0.25)3_14_9_5;POLYGON(0,0,1,0.25)3_5;POLYGON(0,0,1,0.25)46_41;POLYGON(0,0,1,0.25)52_51;POLYGON(0,1,0,0.25)12_29_21_14;POLYGON(0,1,0,0.25)17_19_7_12;POLYGON(0,1,0,0.25)17_19_7_12;POLYGON(0,1,0,0.25)17_19_7_12;POLYGON(0,1,0,0.25)27_37_44_39_29;POLYGON(0,1,0,0.25)3_14_9_5;POLYGON(0.25,0.25,0.25,0.5)35_25;POLYGON(0.25,0.25,0.25,0.5)49_42;POLYGON(0.25,0.25,0.25,0.5)42_37_35;POLYGON(0.25,0.25,0.25,0.5)52_51_44_49;POLYGON(0.25,0.25,0.25,0.5)46_41_31_39;POLYGON(0.25,0.25,0.25,0.5)31_23_16_21;POLYGON(0.625,0.625,0.625,0.5)49_44_37_42;POLYGON(0.625,0.625,0.625,0.5)51_46_39_44;POLYGON(0.625,0.625,0.625,0.5)39_31_21_29;POLYGON(0.625,0.625,0.625,0.5)21_16_9_14;POLYGON(0.625,0.625,0.625,0.5)41_33_23_31;POLYGON(1,0,0,0.25)25_27_19_17;POLYGON(1,0,0,0.25)25_27_19_17;POLYGON(1,0,0,0.25)25_27_19_17;POLYGON(1,0,0,0.25)12_29_21_14;POLYGON(1,0,0,0.25)25_35_37_27;POLYGON(1,0,0,0.25)12_14_7_3;POLYGON(1,0,0,0.25)3_14_9_5;POLYGON(1,0,0,0.25)16_9;POLYGON(1,0,0,0.25)33_23;TICK;MPP_X17*X19*X7*X12;TICK;MPP_X12*X19*X27*X29*X21*X14;TICK;MPP_X27*X37*X44*X39*X29_Z12*Z14*Z7*Z3;TICK;MPP_Z25*Z35*Z37*Z27*Z19*Z17;TICK;MPP_X25*X27*X19*X17;TICK;MPP_Z17*Z19*Z7*Z12;TICK;MPP_Z27*Z29*Z19*Z12;TICK;MPP_X25*X35_X42*X49_X23*X33_X16*X9_Z41*Z46_Z51*Z52_Z5*Z3_Z29*Z39*Z31*Z21;TICK;MPP_Z44*Z51*Z46*Z39_X21*X31*X23*X16_X35*X42*X37_X3*X14*X9*X5;TICK;MPP_Z14*Z21*Z16*Z9_Z31*Z41*Z33*Z23_Z42*Z49*Z44*Z37;TICK;MPP_X49*X52*X51*X44_X39*X46*X41*X31;TICK;R_43_20_8_2_45_30_15_32_47_53_28_13_4;RX_34_26_11_48_36_50_38_40_22_10_24_18_6_0;MARKZ(0)5_9_16_23_33;TICK;CX_18_19_27_28_12_13_6_7_3_4_0_1;TICK;CX_28_27_13_12_7_6_4_3_1_0;TICK;CX_34_25_26_27_11_12_48_42_36_37_19_20_7_8_1_2_38_39_10_9_24_23_46_47_52_53;TICK;CX_34_35_26_25_48_49_27_28_12_13_20_21_8_9_10_16_24_33_51_53_41_47;TICK;CX_26_17_36_42_27_18_12_6_37_28_19_13_7_4_38_29_20_14_8_5;TICK;CX_11_17_36_35_12_18_49_43_19_28_50_52_38_44_20_29_8_14_51_45_39_30_21_15_40_46_22_31_41_32;TICK;CX_25_26_18_19_6_7_28_27_13_12_50_49_44_43_38_37_46_45_40_39_31_30_22_21_16_15_33_32;TICK;CX_17_26_42_43_18_27_6_12_37_28_13_19_4_7_50_51_44_45_29_30_14_15_40_41_31_32_22_23;TICK;CX_35_26_17_11_18_12_28_19_13_7_4_1_29_20_14_8_5_2;TICK;CX_26_27_11_12_37_43_19_20_7_8_1_2_50_44_39_45_21_30_9_15_40_31_22_16_23_32;TICK;CX_28_27_19_18_13_12_7_6_4_3_1_0;TICK;CX_27_28_18_19_12_13_6_7_3_4_0_1;TICK;M_43_20_8_2_45_30_15_32_47_53_28_13_4;MX_34_26_11_48_36_50_38_40_22_10_24_18_6_0;DT(7,2,0)rec[-27]_rec[-30];DT(5,2,0)rec[-26]_rec[-45];DT(3,2,0)rec[-25]_rec[-49];DT(1,4,0)rec[-24]_rec[-38];DT(7,4,0)rec[-23]_rec[-36];DT(5,4,0)rec[-22]_rec[-37];DT(3,4,0)rec[-21]_rec[-32];DT(5,6,0)rec[-20]_rec[-31];DT(6,7,0)rec[-19]_rec[-40];DT(8,5,0)rec[-18]_rec[-39];DT(5,0,0)rec[-17]_rec[-48];DT(4,1,0)rec[-16]_rec[-46];DT(1,3,0)rec[-15];DT(6,1,0)rec[-14]_rec[-44];DT(4,3,0)rec[-13]_rec[-47];DT(2,3,0)rec[-12]_rec[-52];DT(8,3,0)rec[-11]_rec[-43];DT(6,3,0)rec[-10]_rec[-34];DT(9,4,0)rec[-9]_rec[-29];DT(6,5,0)rec[-8]_rec[-50];DT(7,6,0)rec[-7]_rec[-28];DT(4,5,0)rec[-6]_rec[-35];DT(3,6,0)rec[-5]_rec[-41];DT(4,7,0)rec[-4]_rec[-42];DT(4,2,0)rec[-3]_rec[-51];DT(1,2,0)rec[-2]_rec[-33];DT(0,2,0)rec[-1];TICK;MPP_X27*X37*X44*X39*X29_Z17*Z19*Z7*Z12_X3*X14*X9*X5;DT(4,2,1)rec[-3]_rec[-6]_rec[-11];DT(4,1,1)rec[-2];DT(0,2,1)rec[-1]_rec[-4];TICK;MPP_X25*X27*X19*X17_Z12*Z14*Z7*Z3;DT(4,2,2)rec[-2]_rec[-8]_rec[-17]_rec[-18];DT(3,3,2)rec[-1]_rec[-21]_rec[-30];TICK;MPP_Z25*Z35*Z37*Z27*Z19*Z17_Z5*Z3;DT(5,0,3)rec[-2];DT(0,4,3)rec[-1]_rec[-31];TICK;MPP_X12*X19*X27*X29*X21*X14;DT(2,2,4)rec[-1]_rec[-10]_rec[-11];TICK;MPP_Z27*Z29*Z19*Z12;DT(3,3,5)rec[-1]_rec[-25]_rec[-26]_rec[-35];TICK;MPP_X35*X25_X42*X49_X23*X33_X16*X9_Z52*Z51_Z46*Z41_Z29*Z39*Z31*Z21_X17*X19*X7*X12;DT(6,0,6)rec[-8]_rec[-31];DT(8,2,6)rec[-7]_rec[-28];DT(4,8,6)rec[-6]_rec[-21];DT(2,6,6)rec[-5]_rec[-22];DT(9,5,6)rec[-4]_rec[-35];DT(7,7,6)rec[-3]_rec[-36];DT(5,5,6)rec[-2]_rec[-39];DT(2,2,6)rec[-1]_rec[-19]_rec[-20]_rec[-30];TICK;MPP_X35*X42*X37_X49*X52*X51*X44_X39*X46*X41*X31_Z14*Z21*Z16*Z9;DT(6,2,7)rec[-4]_rec[-31]_rec[-34];DT(8,4,7)rec[-3]_rec[-30];DT(6,6,7)rec[-2]_rec[-28];DT(3,5,7)rec[-1]_rec[-42];TICK;MPP_Z42*Z49*Z44*Z37_X21*X31*X23*X16;DT(7,3,8)rec[-2]_rec[-50];DT(4,6,8)rec[-1]_rec[-29];TICK;MPP_Z44*Z51*Z46*Z39_Z31*Z41*Z33*Z23;DT(7,5,9)rec[-2]_rec[-48];DT(5,7,9)rec[-1]_rec[-45]}{Grafting - idling in new code}
\end{itemize}

For complete circuits we ran for each figure, see 
\begin{itemize}
\item \href{https://algassert.com/crumble#circuit=Q(7,8)0;Q(7,9)1;Q(7,10)2;Q(8,7)3;Q(8,8)4;Q(8,9)5;Q(8,10)6;Q(8,11)7;Q(9,8)8;Q(9,9)9;Q(9,10)10;Q(10,9)11;H_YZ[conjugate]_5;TICK;S[injection]_5;TICK;H_3_0_4_1_2_6_10;H_YZ[conjugate]_5;TICK;I[injection]_5;TICK;CZ_0_1_4_5_2_6;TICK;H_4_8_2;TICK;CZ_4_8;TICK;H_8_5_9;I[echo]_4;TICK;CZ_0_4_8_9_1_5;TICK;H_4_9;TICK;I[echo]_8;TICK;CZ_4_8_5_9;TICK;H_9;I[echo]_4;TICK;CZ_3_4_9_10;TICK;H_4_8_1_9_11;TICK;I[echo]_5;TICK;CZ_0_4_8_9;TICK;I[echo]_1_11;TICK;CZ_4_5_1_2_9_11;TICK;H_1_11_10;I[echo]_4_8_9;TICK;CZ_0_1_4_8_5_9_6_10;TICK;H_0_4_8_9_6_10;I[echo]_1;TICK;CZ_3_4_1_5_9_11_2_6;TICK;H_4_1_5_9_6;TICK;I[echo]_4_9;TICK;I[echo]_8_5_11_2_10_7_3_0;M_4_1_6_9;DT(8,8,0)rec[-4];DT(7,9,0)rec[-3];DT(8,10,0)rec[-2];DT(9,9,0)rec[-1];TICK;R_4_1_9_6;TICK;I[conjugate]_3_0_8_5_11_2_10;TICK;I[echo]_8_5_11_2_7_3_0_10;TICK;I[T_gate]_3_0_8_5_11_2_10;TICK;I[conjugate]_3_0_8_5_11_2_10;TICK;H_0_4_1_5_9_11_2_6;TICK;CZ_0_4_1_5_9_11_2_6;TICK;H_3_5_6_10;I[echo]_4;TICK;CZ_3_4_5_6_9_10;TICK;H_4_8_6_7;I[echo]_5_9;TICK;CZ_4_5_8_9_6_7;TICK;H_5_6_7;TICK;CZ_5_9_6_7;TICK;H_9_6;TICK;I[echo]_4_8_1_5_11_2_10_7_3_0;M_9_6;DT(9,9,1)rec[-2];DT(8,10,1)rec[-1];TICK;R_9_6;TICK;H_9_6;TICK;CZ_5_9_6_7;TICK;H_5_6_7;TICK;CZ_4_5_8_9_6_7;TICK;H_4_8_6_7;I[echo]_5;TICK;CZ_3_4_5_6_9_10;TICK;H_3_5_6_10;TICK;CZ_0_4_1_5_9_11_2_6;TICK;H_0_5_11_2;TICK;I[conjugate]_3_0_8_5_11_2_10;TICK;I[echo]_4_8_1_5_11_6_2_7_3_0_10;TICK;I[T_dagger_gate]_3_0_8_5_11_2_10;TICK;I[conjugate]_3_0_8_5_11_2_10;TICK;H_3_0_4_8_1_5_9_11_2_6_10;TICK;M_4_1_9_6_3_0_8_5_11_2_10_7;DT(8,8,2)rec[-12];DT(7,9,2)rec[-11];DT(9,9,2)rec[-10]_rec[-14];DT(8,10,2)rec[-9];DT(8,9,2)rec[-5]_rec[-6]_rec[-7]_rec[-8];DT(9,10,2)rec[-2]_rec[-3]_rec[-5]_rec[-7];DT(10,9,2)rec[-2]_rec[-4]_rec[-5]_rec[-6];DT(8,11,2)rec[-1];OI(0)rec[-2]_rec[-3]_rec[-4]_rec[-5]_rec[-6]_rec[-7]_rec[-8]_rec[-9]_rec[-10]_rec[-12]}{Figure 2: cultivation, cyan curve}
\item \href{https://algassert.com/crumble#circuit=Q(3,3)0;Q(3,4)1;Q(3,5)2;Q(3,6)3;Q(3,7)4;Q(4,3)5;Q(4,4)6;Q(4,5)7;Q(4,6)8;Q(4,7)9;Q(5,4)10;Q(5,5)11;Q(5,6)12;Q(6,5)13;H_YZ[conjugate]_7;TICK;S[injection]_7;TICK;H_5_1_6_2_3_8_12;H_YZ[conjugate]_7;TICK;I[injection]_7;TICK;CZ_1_2_6_7_3_8;TICK;H_6_10_3;TICK;CZ_6_10;TICK;H_10_7_11;I[echo]_6;TICK;CZ_1_6_10_11_2_7;TICK;H_6_11;TICK;I[echo]_10;TICK;CZ_6_10_7_11;TICK;H_11;I[echo]_6;TICK;CZ_5_6_11_12;TICK;H_6_10_2_11_13;TICK;I[echo]_7;TICK;CZ_1_6_10_11;TICK;I[echo]_2_13;TICK;CZ_6_7_2_3_11_13;TICK;H_2_13_12;I[echo]_6_10_11;TICK;CZ_1_2_6_10_7_11_8_12;TICK;H_1_6_10_11_8_12;I[echo]_2;TICK;CZ_5_6_2_7_11_13_3_8;TICK;H_6_2_7_11_8;TICK;I[echo]_6_11;TICK;I[echo]_5_1_10_7_13_3_12_9;M_6_2_8_11;DT(4,4,0)rec[-4];DT(3,5,0)rec[-3];DT(4,6,0)rec[-2];DT(5,5,0)rec[-1];TICK;R_6_2_11_8;TICK;H_0_6_2_11_8_4;TICK;CZ_0_5_3_4_8_12;TICK;H_0_5_3_8_12_4;I[echo]_6_2_11;TICK;CZ_5_6_2_3_11_12;TICK;H_5_6_2_11_12;TICK;I[echo]_3_8;TICK;CZ_6_10_2_7_11_13_3_8;TICK;I[echo]_6_2_11;TICK;CZ_1_2_6_7_10_11;TICK;CZ_1_6_7_11;TICK;H_1_6_10_2_7_11_13_3_8;TICK;I[echo]_4_0_5_12;TICK;CZ_1_6_7_11;TICK;I[echo]_1_6_10_2_7_11_8_13;TICK;CZ_1_2_6_7_10_11;TICK;I[echo]_3;TICK;CZ_6_10_2_7_11_13_3_8;TICK;H_1_6_10_2_7_11_13_8;TICK;CZ_5_6_2_3_11_12;TICK;H_5_6_2_11_3_12;I[echo]_8;TICK;CZ_0_5_3_4_8_12;TICK;H_0_5_3_8_12_4;TICK;I[echo]_6_8;TICK;I[echo]_5_1_10_7_13_3_12_9;M_6_2_11_0_8_4;DT(4,4,1)rec[-6];DT(3,5,1)rec[-5];DT(5,5,1)rec[-4];DT(3,3,1)rec[-3];DT(4,6,1)rec[-2];DT(3,7,1)rec[-1];TICK;R_6_2_11_8_4_0;TICK;I[conjugate]_5_1_10_7_13_3_12;TICK;I[echo]_5_1_10_7_13_3;TICK;I[T_gate]_5_1_10_7_13_3_12;TICK;I[conjugate]_5_1_10_7_13_3_12;TICK;H_1_6_2_7_11_13_3_8;TICK;CZ_1_6_2_7_11_13_3_8;TICK;H_5_7_8_12;I[echo]_6;TICK;CZ_5_6_7_8_11_12;TICK;H_6_10_8_9;I[echo]_7_11;TICK;CZ_6_7_10_11_8_9;TICK;H_7_8_9;TICK;CZ_7_11_8_9;TICK;H_11_8;TICK;I[echo]_5_1_6_10_2_7_13_3_12_9;M_11_8;DT(3,7,2)rec[-2]_rec[-3]_rec[-4]_rec[-5];DT(4,6,2)rec[-1];TICK;R_11_8;TICK;H_11_8;TICK;CZ_7_11_8_9;TICK;H_7_8_9;TICK;CZ_6_7_10_11_8_9;TICK;H_6_10_8;I[echo]_7_11;TICK;CZ_5_6_7_8_11_12;TICK;H_5_7_8_12;I[echo]_6;TICK;CZ_1_6_2_7_11_13_3_8;TICK;H_1_6_2_7_11_13_3_8;TICK;I[conjugate]_5_1_10_7_13_3_12;TICK;I[echo]_5_1_6_10_2_7_11_13_8_3_9_12;TICK;I[T_dagger_gate]_5_1_10_7_13_3_12;TICK;I[conjugate]_5_1_10_7_13_3_12;TICK;I[echo]_5_1_10_7_13_3_12_9;M_6_2_11_8;DT(4,4,3)rec[-4];DT(3,5,3)rec[-3];DT(5,5,3)rec[-2]_rec[-6];DT(4,6,3)rec[-1];TICK;R_0_6_2_4_11_8;TICK;H_0_6_2_11_8_4;TICK;CZ_0_5_3_4_8_12;TICK;H_0_5_3_8_12_4;I[echo]_6_2_11;TICK;CZ_5_6_2_3_11_12;TICK;H_5_6_2_11_12;TICK;I[echo]_3_8;TICK;CZ_6_10_2_7_11_13_3_8;TICK;I[echo]_6_2_11;TICK;CZ_1_2_6_7_10_11;TICK;CZ_1_6_7_11;TICK;H_1_6_10_2_7_11_13_3_8;TICK;I[echo]_4_0_5_12;TICK;CZ_1_6_7_11;TICK;I[echo]_1_6_10_2_7_11_13_8;TICK;CZ_1_2_6_7_10_11;TICK;I[echo]_3;TICK;CZ_6_10_2_7_11_13_3_8;TICK;H_1_6_10_2_7_11_13_8;TICK;CZ_5_6_2_3_11_12;TICK;H_5_6_2_11_3_12;I[echo]_8;TICK;CZ_0_5_3_4_8_12;TICK;H_0_5_3_8_12_4;TICK;I[echo]_2;TICK;I[echo]_5_1_10_7_13_3_12_9;M_6_2_11_0_8_4;DT(4,4,4)rec[-6];DT(5,5,4)rec[-5]_rec[-11]_rec[-16]_rec[-17];DT(3,5,4)rec[-4]_rec[-17];DT(3,3,4)rec[-3];DT(4,6,4)rec[-2];DT(3,7,4)rec[-1]_rec[-14];TICK;R_6_2_11_8_4_0;TICK;I[conjugate]_5_1_10_7_13_3_12;TICK;I[echo]_5_1_7_13_3;TICK;I[T_gate]_5_1_10_7_13_3_12;TICK;I[conjugate]_5_1_10_7_13_3_12;TICK;H_1_6_2_7_11_13_3_8;TICK;CZ_1_6_2_7_11_13_3_8;TICK;H_5_7_8_12;I[echo]_6;TICK;CZ_5_6_7_8_11_12;TICK;H_6_10_8;I[echo]_7_11;TICK;CZ_6_7_10_11_8_9;TICK;H_7_8_9;TICK;CZ_7_11_8_9;TICK;H_11_8;TICK;I[echo]_5_1_6_10_2_7_13_3_12_9;M_11_8;DT(4,6,5)rec[-1];OI(0)rec[-2]_rec[-3]_rec[-4]_rec[-5]_rec[-9]_rec[-10]_rec[-12];TICK;R_11_8;TICK;H_11_8;TICK;CZ_7_11_8_9;TICK;H_7_8_9;TICK;CZ_6_7_10_11_8_9;TICK;H_6_10_8;I[echo]_7_11;TICK;CZ_5_6_7_8_11_12;TICK;H_5_7_8_12;I[echo]_6;TICK;CZ_1_6_2_7_11_13_3_8;TICK;H_1_6_2_7_11_13_3_8;TICK;I[conjugate]_5_1_10_7_13_3_12;TICK;I[echo]_5_1_6_10_2_7_11_13_8_3_9_12;TICK;I[T_dagger_gate]_5_1_10_7_13_3_12;TICK;I[conjugate]_5_1_10_7_13_3_12;TICK;I[echo]_5_1_10_7_13_3_12_9;M_6_2_11_8;DT(4,4,6)rec[-4];DT(3,5,6)rec[-3];DT(5,5,6)rec[-2]_rec[-6];DT(4,6,6)rec[-1];TICK;R_0_6_2_4_11_8;TICK;H_0_6_2_11_8_4;TICK;CZ_0_5_3_4_8_12;TICK;H_0_5_3_8_12_4;I[echo]_6_2_11;TICK;CZ_5_6_2_3_11_12;TICK;H_5_6_2_11_12;TICK;I[echo]_3_8;TICK;CZ_6_10_2_7_11_13_3_8;TICK;I[echo]_6_2_11;TICK;CZ_1_2_6_7_10_11;TICK;CZ_1_6_7_11;TICK;H_1_6_10_2_7_11_13_3_8;TICK;I[echo]_4_0_5_12;TICK;CZ_1_6_7_11;TICK;I[echo]_1_6_10_2_7_11_13_8;TICK;CZ_1_2_6_7_10_11;TICK;I[echo]_3;TICK;CZ_6_10_2_7_11_13_3_8;TICK;H_1_6_10_2_7_11_13_8;TICK;CZ_5_6_2_3_11_12;TICK;H_5_6_2_11_3_12;I[echo]_8;TICK;CZ_0_5_3_4_8_12;TICK;H_0_5_3_8_12_4;TICK;I[echo]_2;TICK;I[echo]_5_1_10_7_13_3_12_9;M_6_2_11_0_8_4;DT(4,4,7)rec[-6];DT(5,5,7)rec[-5]_rec[-11]_rec[-16]_rec[-17];DT(3,5,7)rec[-4]_rec[-17];DT(3,3,7)rec[-3];DT(4,6,7)rec[-2];DT(3,7,7)rec[-1]_rec[-14];TICK;R_6_2_11_8_4_0;TICK;I[conjugate]_5_1_10_7_13_3_12;TICK;I[echo]_5_1_7_13_3;TICK;I[T_gate]_5_1_10_7_13_3_12;TICK;I[conjugate]_5_1_10_7_13_3_12;TICK;H_1_6_2_7_11_13_3_8;TICK;CZ_1_6_2_7_11_13_3_8;TICK;H_5_7_8_12;I[echo]_6;TICK;CZ_5_6_7_8_11_12;TICK;H_6_10_8;I[echo]_7_11;TICK;CZ_6_7_10_11_8_9;TICK;H_7_8_9;TICK;CZ_7_11_8_9;TICK;H_11_8;TICK;I[echo]_5_1_6_10_2_7_13_3_12_9;M_11_8;DT(4,6,8)rec[-2]_rec[-3]_rec[-4]_rec[-5]_rec[-9]_rec[-10]_rec[-12];DT(4,6,9)rec[-1];TICK;R_11_8;TICK;H_11_8;TICK;CZ_7_11_8_9;TICK;H_7_8_9;TICK;CZ_6_7_10_11_8_9;TICK;H_6_10_8_9;I[echo]_7_11;TICK;CZ_5_6_7_8_11_12;TICK;H_5_7_8_12;I[echo]_6;TICK;CZ_1_6_2_7_11_13_3_8;TICK;H_1_6_2_7_11_13_3_8;TICK;I[conjugate]_5_1_10_7_13_3_12;TICK;I[echo]_5_1_6_10_2_7_11_13_8_3_9_12;TICK;I[T_dagger_gate]_5_1_10_7_13_3_12;TICK;I[conjugate]_5_1_10_7_13_3_12;TICK;I[echo]_5_1_10_7_13_3_12_9;M_6_2_11_8;DT(4,4,10)rec[-4];DT(3,5,10)rec[-3];DT(5,5,10)rec[-2]_rec[-6];DT(4,6,10)rec[-1];TICK;R_0_6_2_4_11_8;TICK;H_0_6_2_11_8_4;TICK;CZ_0_5_3_4_8_12;TICK;H_0_5_3_8_12_4;I[echo]_6_2_11;TICK;CZ_5_6_2_3_11_12;TICK;H_5_6_2_11_12;TICK;I[echo]_3_8;TICK;CZ_6_10_2_7_11_13_3_8;TICK;I[echo]_6_2_11;TICK;CZ_1_2_6_7_10_11;TICK;CZ_1_6_7_11;TICK;H_1_6_10_2_7_11_13_3_8;TICK;I[echo]_4_0_5_12;TICK;CZ_1_6_7_11;TICK;I[echo]_6_2_11_8;TICK;CZ_1_2_6_7_10_11;TICK;I[echo]_3;TICK;CZ_6_10_2_7_11_13_3_8;TICK;H_6_2_11_8;TICK;CZ_5_6_2_3_11_12;TICK;H_5_6_2_11_3_12;I[echo]_8;TICK;CZ_0_5_3_4_8_12;TICK;H_0_8_4;TICK;I[echo]_2_1_12_9;TICK;M_6_2_11_0_8_4_5_1_10_7_13_3_12_9;DT(4,4,11)rec[-14];DT(5,5,11)rec[-13]_rec[-19]_rec[-24]_rec[-25];DT(3,5,11)rec[-12]_rec[-25];DT(3,3,11)rec[-11];DT(4,6,11)rec[-10];DT(3,7,11)rec[-9]_rec[-22];DT(4,5,11)rec[-5]_rec[-6]_rec[-7]_rec[-8];DT(5,6,11)rec[-2]_rec[-3]_rec[-5]_rec[-7]_rec[-10];DT(6,5,11)rec[-2]_rec[-4]_rec[-5]_rec[-6];DT(4,7,11)rec[-1]}{Figure 3: kickback tomography, cyan curve}
\item \href{https://algassert.com/crumble#circuit=Q(3,7)0;Q(3,8)1;Q(4,5)2;Q(4,6)3;Q(4,7)4;Q(4,8)5;Q(5,2)6;Q(5,3)7;Q(5,4)8;Q(5,5)9;Q(5,6)10;Q(5,7)11;Q(5,8)12;Q(5,9)13;Q(5,10)14;Q(6,2)15;Q(6,3)16;Q(6,4)17;Q(6,5)18;Q(6,6)19;Q(6,7)20;Q(6,8)21;Q(6,9)22;Q(6,10)23;Q(7,2)24;Q(7,3)25;Q(7,4)26;Q(7,5)27;Q(7,6)28;Q(7,7)29;Q(7,8)30;Q(7,9)31;Q(7,10)32;Q(7,11)33;Q(8,4)34;Q(8,5)35;Q(8,6)36;Q(8,7)37;Q(8,8)38;Q(8,9)39;Q(8,10)40;Q(8,11)41;Q(9,4)42;Q(9,5)43;Q(9,6)44;Q(9,7)45;Q(9,8)46;Q(9,9)47;Q(10,6)48;Q(10,7)49;Q(10,8)50;Q(10,9)51;Q(11,6)52;Q(11,7)53;H_YZ[conjugate]_10;TICK;S[injection]_10;TICK;H_8_2_9_3_4_11_20;H_YZ[conjugate]_10;TICK;I[injection]_10;TICK;CZ_2_3_9_10_4_11;TICK;H_9_18_4;TICK;CZ_9_18;TICK;H_18_10_19;I[echo]_9;TICK;CZ_2_9_18_19_3_10;TICK;H_9_19;TICK;I[echo]_18;TICK;CZ_9_18_10_19;TICK;H_19;I[echo]_9;TICK;CZ_8_9_19_20;TICK;H_9_18_3_19_28;TICK;I[echo]_10;TICK;CZ_2_9_18_19;TICK;I[echo]_28_3;TICK;CZ_9_10_3_4_19_28;TICK;H_3_28_20;I[echo]_9_18_19;TICK;CZ_2_3_9_18_10_19_11_20;TICK;H_2_9_18_19_11_20;I[echo]_3;TICK;CZ_8_9_3_10_19_28_4_11;TICK;H_9_3_10_19_11;TICK;I[echo]_9_19;TICK;I[echo]_8_10_28_12_2_18_4_20;M_9_3_11_19;DT(5,5,0)rec[-4];DT(4,6,0)rec[-3];DT(5,7,0)rec[-2];DT(6,6,0)rec[-1];TICK;R_9_3_19_11;TICK;I[conjugate]_8_2_18_10_28_4_20;TICK;I[echo]_8_2_18_10_28_4_12_20;TICK;I[T_gate]_8_2_18_10_28_4_20;TICK;I[conjugate]_8_2_18_10_28_4_20;TICK;H_2_9_3_10_19_28_4_11;TICK;CZ_2_9_3_10_19_28_4_11;TICK;H_8_10_11_20;I[echo]_9;TICK;CZ_8_9_10_11_19_20;TICK;H_9_18_11_12;I[echo]_10_19;TICK;CZ_9_10_18_19_11_12;TICK;H_10_11_12;TICK;CZ_10_19_11_12;TICK;H_19_11;TICK;I[echo]_8_3_10_28_12_2_9_18_4_20;M_19_11;DT(6,6,1)rec[-2];DT(5,7,1)rec[-1];TICK;R_19_11;TICK;H_19_11;TICK;CZ_10_19_11_12;TICK;H_10_11_12;TICK;CZ_9_10_18_19_11_12;TICK;H_9_18_11;I[echo]_10_19;TICK;CZ_8_9_10_11_19_20;TICK;H_8_10_11_20;I[echo]_9;TICK;CZ_2_9_3_10_19_28_4_11;TICK;H_2_9_3_10_19_28_4_11;TICK;I[conjugate]_8_2_18_10_28_4_20;TICK;I[echo]_8_2_9_18_3_10_19_28_11_4_12_20;TICK;I[T_dagger_gate]_8_2_18_10_28_4_20;TICK;I[conjugate]_8_2_18_10_28_4_20;TICK;I[echo]_8_10_28_2_18_4_20;M_9_3_19_11;DT(5,5,2)rec[-4];DT(4,6,2)rec[-3];DT(6,6,2)rec[-2]_rec[-6];DT(5,7,2)rec[-1];TICK;R_11_9_12_3_19;TICK;H_6_24_7_25_17_26_34_2_9_27_35_43_3_10_19_36_48_0_4_11_29_37_45_49_53_1_5_12_21_30_38_46_50_13_31_47_14_23_32_40_33;TICK;CZ_6_15_24_25_26_27_34_35_42_43_2_3_10_19_28_29_36_37_44_45_48_49_52_53_0_4_1_5_12_21_30_38_46_47_50_51_13_22_31_39_14_23_32_40_33_41;TICK;H_15_24_25_26_34_42_2_27_35_43_3_28_36_44_48_52_0_4_29_37_45_49_53_1_5_12_21_30_38_46_50_13_22_31_39_47_51_14_23_32_40_33_41;TICK;I[echo]_9_11;TICK;CZ_24_25_26_27_34_35_42_43_2_3_28_29_36_37_44_45_48_49_52_53_0_4_1_5_12_21_30_38_46_47_50_51_13_22_31_39_14_23_32_40_33_41;TICK;H_34_48_13_47_40;I[echo]_26_28_36_21_38_25_27_29;TICK;CZ_15_24_17_26_34_42_2_9_19_28_48_52_0_1_4_11_12_21_30_38_13_14_47_51_33_41;TICK;H_25_8_26_9_18_27_28_36_11_20_29_21_38;I[echo]_17_10_19_23_43_0_53_31_51_33_41;TICK;I[echo]_34_3_5_4_45_13_47;TICK;CZ_26_34_42_43_9_18_28_36_52_53_0_4_11_20_1_5_50_51_14_23_40_41;TICK;I[echo]_8_36_50_25_18;TICK;CZ_16_17_25_26_8_9_18_19_27_28_3_4_10_11_20_21_29_30_12_13;TICK;H_16_25_8_34_18_43_36_0_53_13_31_47_51_23_33_41;I[echo]_6_15_21_38_40_2_9_27_11_20_29;TICK;I[echo]_48_7;TICK;CZ_26_27_2_3_9_10_35_36_19_20_28_29_44_45_37_38_49_50_5_12_30_31_46_47_22_23_39_40_32_33;TICK;H_17_9_27_3_10_19_36_4_11_20_29_45_5_21_38_31_47_23_33;I[echo]_24_26_28_35_37_49_22_39;TICK;I[echo]_42_1_12_30_46_14_32_44_52;TICK;CZ_8_17_9_18_35_43_10_19_36_44_0_4_11_20_37_45_49_53_21_30_38_46_22_31_39_47_23_32;TICK;H_36_20_21_38_50_40;I[echo]_8_43;TICK;I[echo]_17_10_19_4_45_31;TICK;CZ_16_17_8_9_18_19_27_35_3_4_10_11_44_48_20_21_29_37_45_49_46_50_13_22_31_39_32_40;TICK;H_15_26_2_28_48_20_50_40;I[echo]_16_18;TICK;I[echo]_9_11;TICK;CZ_15_16_24_25_17_18_26_27_2_3_9_10_19_20_28_29_4_5;TICK;H_6_15_8_17_2_43_10_19_4_45_31;TICK;I[echo]_26_28_20;TICK;CZ_15_24_17_26_34_35_2_9_43_44_19_28_36_37_48_49_4_11_45_46_21_22_38_39_31_32;TICK;I[echo]_6_15_8_17_10_19;TICK;CZ_6_15_7_16_8_17_9_18_10_19_11_20;TICK;H_6_15_24_7_16_8_17_26_9_18_35_10_19_28_11_20_37_49_22_39;I[echo]_2_4;TICK;CZ_6_15_7_16_8_17_9_18_10_19_11_20;TICK;H_6_16_8_42_2_18_10_44_52_4_20_1_12_30_46_14_32;TICK;I[echo]_22_37_49_4_14_12_46_44_42_52_10_6_8;TICK;I[echo]_15_17_34_3_19_36_48_21_5_50_23_40_38_7_25_9_27_0_11_43_29_45_13_31_53_47_33;M_22_28_26_24_39_37_35_49_51_41_20_18_16_1_4_2_14_12_32_30_46_44_42_52_10_6_8;DT(6,9,3)rec[-27];DT(7,6,3)rec[-26];DT(8,9,3)rec[-23];DT(10,9,3)rec[-19];DT(8,11,3)rec[-18];DT(5,7,3)rec[-17]_rec[-32];DT(6,3,3)rec[-15]_rec[-16];DT(6,3,4)rec[-15];DT(4,7,3)rec[-13];DT(4,5,3)rec[-12];DT(9,6,3)rec[-6];DT(9,4,3)rec[-5];DT(11,6,3)rec[-4];DT(5,6,3)rec[-3];DT(5,2,3)rec[-2];DT(5,4,3)rec[-1];TICK;R_22_28_26_24_39_37_35_49_51_41_20_18_16_1_4_2_14_12_32_30_46_44_42_52_10_8_6;TICK;H_6_24_16_8_26_42_2_18_35_10_28_44_52_4_20_37_49_1_12_30_46_22_39_51_14_32_41;TICK;CZ_6_15_7_16_8_17_9_18_10_19_11_20;TICK;H_6_15_7_16_8_17_34_9_18_19_48_0_11_20_21_38_13;TICK;I[echo]_26_10_28;TICK;CZ_6_15_7_16_8_17_9_18_11_20;TICK;I[echo]_21_38_16_9_18_11_20_51_41;TICK;CZ_15_24_17_26_34_42_2_9_19_28_48_52_0_1_4_11_12_21_30_38_13_14_47_51_33_41;TICK;H_16_25_26_9_18_27_3_10_28_36_11_20_29_53_5_21_38;I[echo]_8_34_4_0_13;TICK;I[echo]_17_19;TICK;CZ_26_34_42_43_9_18_28_36_52_53_0_4_11_20_1_5_50_51_14_23_40_41;TICK;I[echo]_36_16_25_2_18_53;TICK;CZ_16_17_25_26_8_9_18_19_27_28_3_4_10_11_20_21_29_30_12_13;TICK;H_16_25_8_34_18_43_36_0_53_13_47_51_23_33_41;I[echo]_15_3_10_48_5_21_38_9_27_11_20_29;TICK;CZ_26_27_2_3_9_10_35_36_19_20_28_29_44_45_37_38_49_50_5_12_30_31_46_47_22_23_39_40_32_33;TICK;H_17_9_27_3_10_19_36_4_11_20_29_45_5_21_38_31_47_23;I[echo]_24_26_28_35_37_49_22_39;TICK;I[echo]_6_42_1_12_30_46_14_32_7_44_52;TICK;CZ_8_17_9_18_35_43_10_19_36_44_0_4_11_20_37_45_49_53_21_30_38_46_22_31_39_47_23_32;TICK;H_36_20_21_38_50_40;I[echo]_8_43;TICK;I[echo]_17_10_19_4_45_31;TICK;CZ_16_17_8_9_18_19_27_35_3_4_10_11_44_48_20_21_29_37_45_49_46_50_13_22_31_39_32_40;TICK;H_15_26_2_28_48_20;I[echo]_16_18;TICK;I[echo]_34_25_9_11_0_53_13;TICK;CZ_15_16_24_25_17_18_26_27_2_3_9_10_19_20_28_29_4_5;TICK;H_15_8_17_2_43_10_19_4_45_31;I[echo]_3_5_27_29_47;TICK;I[echo]_26_28_36_21_38_20;TICK;CZ_15_24_17_26_34_35_2_9_43_44_19_28_36_37_48_49_4_11_45_46_21_22_38_39_31_32;TICK;I[echo]_15_8_17_10_19_48;TICK;CZ_6_15_7_16_8_17_9_18_10_19_11_20;TICK;H_6_15_24_7_16_8_17_26_9_18_35_10_19_28_11_20_37_49_22_39;I[echo]_2_4;TICK;CZ_6_15_7_16_8_17_9_18_10_19_11_20;TICK;H_6_7_16_25_8_34_42_2_9_18_27_3_10_36_44_48_52_0_4_11_20_29_53_1_5_12_21_30_38_46_13_47_14_32;TICK;I[echo]_22_28_26_37_49_51_41_20_18_4_14_30_6_15_25_17_34_9_36_48_11_53_5_21_38_50_13_31_40;TICK;M_22_28_26_24_39_37_35_49_51_41_20_18_16_1_4_2_14_12_32_30_46_44_42_52_10_8_6_15_7_25_17_34_9_27_43_3_19_36_48_0_11_29_45_53_5_21_38_50_13_31_47_23_40_33;DT(6,9,5)rec[-54]_rec[-81];DT(6,5,5)rec[-53]_rec[-70]_rec[-71]_rec[-80];DT(7,4,5)rec[-52]_rec[-70]_rec[-79];DT(6,3,5)rec[-51]_rec[-69]_rec[-78];DT(8,11,5)rec[-50]_rec[-72]_rec[-77];DT(8,7,5)rec[-49]_rec[-76];DT(8,5,5)rec[-48]_rec[-75];DT(10,7,5)rec[-47]_rec[-74];DT(10,9,5)rec[-46]_rec[-73];DT(8,11,6)rec[-45]_rec[-72];DT(6,7,5)rec[-44];DT(6,5,6)rec[-43];DT(6,3,6)rec[-42];DT(3,8,5)rec[-41]_rec[-68];DT(5,6,5)rec[-40]_rec[-57]_rec[-66]_rec[-67];DT(5,4,5)rec[-39]_rec[-55]_rec[-57]_rec[-67];DT(5,10,5)rec[-38]_rec[-65];DT(5,8,5)rec[-37]_rec[-64]_rec[-67];DT(7,10,5)rec[-36]_rec[-63];DT(7,8,5)rec[-35]_rec[-57]_rec[-62];DT(9,8,5)rec[-34]_rec[-61];DT(9,6,5)rec[-33]_rec[-60];DT(9,4,5)rec[-32]_rec[-59];DT(11,6,5)rec[-31]_rec[-58];DT(5,6,6)rec[-30]_rec[-57];DT(5,2,5)rec[-29]_rec[-56];DT(5,2,6)rec[-28];DT(6,2,5)rec[-27];DT(7,5,5)rec[-21]_rec[-23]_rec[-25]_rec[-26]_rec[-28];DT(9,5,5)rec[-20]_rec[-23]_rec[-32];DT(6,6,5)rec[-18]_rec[-19]_rec[-22]_rec[-24]_rec[-29]_rec[-30]_rec[-40];DT(5,7,5)rec[-14]_rec[-15]_rec[-18]_rec[-19]_rec[-30]_rec[-39]_rec[-40];DT(7,7,5)rec[-13]_rec[-14]_rec[-17]_rec[-18]_rec[-21]_rec[-22]_rec[-29]_rec[-30];DT(9,7,5)rec[-12]_rec[-16]_rec[-17]_rec[-20]_rec[-33];DT(11,7,5)rec[-11]_rec[-16]_rec[-31];DT(4,8,5)rec[-10]_rec[-15]_rec[-41];DT(5,9,5)rec[-6]_rec[-9]_rec[-10]_rec[-37]_rec[-40];DT(7,9,5)rec[-5]_rec[-8]_rec[-9]_rec[-13]_rec[-14]_rec[-30]_rec[-35];DT(9,9,5)rec[-4]_rec[-7]_rec[-8]_rec[-12]_rec[-34];DT(6,10,5)rec[-3]_rec[-6]_rec[-38];DT(7,11,5)rec[-1]_rec[-2]_rec[-3]_rec[-5]_rec[-36];OI(0)rec[-15]_rec[-19]_rec[-24]_rec[-25]_rec[-26]_rec[-29]_rec[-39]_rec[-57]_rec[-66]_rec[-82]_rec[-83]_rec[-85]}{Figure 4: grafting, $N = 2$, with cultivation, cyan curve}
\end{itemize}

\section{Reinforcement learning calibration}
\begin{figure}
    \centering
\includegraphics[width=0.5\linewidth]{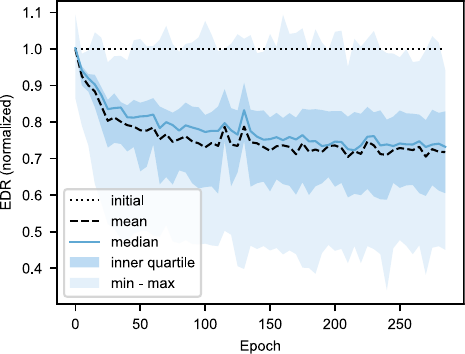}
    \caption{Normalized error detection rate (EDR) as a function of the training epoch in the RL process. Normalization is done by computing the ratio of EDR achieved by the RL agent to the EDR at epoch 0 achieved with our traditional calibration stack.}\label{fig:rlc_training}
\end{figure}

We use reinforcement learning (RL) control framework introduced in Ref.~\cite{sivak_reinforcement_2025} as a final step in our calibration protocol. This framework tunes the control parameters with an objective to minimize the rate of error detection events. The optimized control parameters are the same as in Supplementary Material Section II of Ref.\cite{sivak_reinforcement_2025}. For the kickback tomography data in Fig. 3 of the main text where an error of $\sim 1 \times 10^{-4}$ is observed, we find an improvement in the error detection rate (EDR) by about 30\% on average, as shown in Fig.~\ref{fig:rlc_training}. Training occurs over about 4 hours immediately before the main text data is taken. 
\begin{figure}
    \centering
\includegraphics[width=0.3\linewidth]{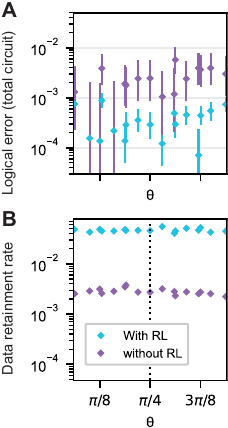}
    \caption{Logical error probability of the full circuit as measured via kickback tomography, and data retainment rate, with and without RL calibration. In our experiment, RL calibration was crucial for observing the magic state fidelity and post selection rates  reported in the main text.}
    \label{fig:with_without_RL}
\end{figure}
In Fig.~\ref{fig:with_without_RL}, we show the difference in performance when adding the RL control technique on top of the traditional calibration stack. We compare the logical error probability for the full circuit using kickback tomography, as well as the data retainment rates. Both metrics are improved by about an order of magnitude by the RL agent.

\section{\label{sec:nonphysical_tomo} Unphysical expectations and post selected logical tomography}

In this section, we discuss the main reason we do not report a state fidelity using the data in Fig. 2 of the main text: post selected logical state tomography may overestimate fidelity in the presence of coherent noise. 

The qubit state output by a process can be estimated by tomography; by separately sampling $\langle X \rangle$, $\langle Y \rangle$, and $\langle Z \rangle$ in order to derive a Bloch vector $(\langle X \rangle, \langle Y \rangle, \langle Z \rangle)$.
A well known downside of tomography is that the estimated Bloch vector can end up outside the Bloch sphere ($\langle X \rangle^2 + \langle Y \rangle^2 + \langle Z \rangle^2 > 1$); it can produce an estimate that doesn't correspond to a physical state.

In the ideal case, a non-physical estimate would be due to shot noise.
In that case, the physical state could be resolved by simply taking more shots.
Unfortunately, there are also more insidious causes of apparent non-physicality.
For example, it's conceivable that calibration drift could happen to bias the state slightly towards the X axis while estimating $\langle X \rangle$ and slightly towards the Y axis while estimating $\langle Y \rangle$ resulting in $\langle X \rangle^2 + \langle Y \rangle^2 > 1$.
Taking more shots may not resolve this issue (a better resolution in this case would be to interleave the shots being used for each of the three coordinate estimates).
Another example, and the one we will be discussing here in more detail, is the possibility of systemic bias due to non-uniform post selection.

When a state is encoded into a quantum error correcting code, a natural thing to do in order to improve the tomography of that state is to discard shots where the error correcting code has indicated there's a problem, as we do in the main text.
For example, if a logical state is being measured transversally, the measurements will reveal not only the logical measurement result but also the values of some of the stabilizers of the code.
If one of the stabilizers is in the wrong state, this suggests there was some sort of error during the tomography.
A natural thing to do in that situation is to discard the shot.
This especially makes sense in the context of cultivation, which unconditionally discards shots when stabilizer measurements are seen to be wrong (while the system is at a small code distance).
We will refer to doing tomography on the subset of shots where all revealed stabilizers have the expected value as ``post selected logical tomography''.

Consider the seven-qubit quantum state $|T_C\rangle$ that corresponds to a distance-3 color code with its $X$ and $Z$ stabilizer generators in the +1 eigenbasis, and its logical qubit in the state $|T_z \rangle = \frac{1}{2}\left(|0\rangle + e^{i\pi/4}|1\rangle\right)$.
Let $\hat{U}$ be a single qubit operation that corresponds to the following actions: rotate $-10^\circ$ around the $Z$ axis, then $-10^\circ$ around the $X$ axis, then $-10^\circ$ around the $Z$ axis.
Let $|\widetilde{T_C}\rangle = U^{\otimes 7} |T_C\rangle$ be the pure state produced by applying $\hat{U}$ to each of the seven data qubits of the color code state.
If we compute what happens during noiseless post selected logical tomography of $|\widetilde{T_C}\rangle$, we find:

\begin{itemize}
    \item The chance of keeping an $X$ axis tomography shot is $\approx 82.0\%$.
    \item The chance of keeping a $Y$ axis tomography shot is $\approx 94.7\%$.
    \item The chance of keeping a $Z$ axis tomography shot is $\approx 100.0\%$.
    \item Within kept shots, $\langle \hat{X}_L \rangle_\text{kept} \approx 0.6530$.
    \item Within kept shots, $\langle \hat{Y}_L \rangle_\text{kept} \approx -0.7588$.
    \item Within kept shots, $\langle \hat{Z}_L \rangle_\text{kept} \approx 0.0072$.
\end{itemize}

Note that $X_\text{kept}^2 + Y_\text{kept}^2 \approx 1.0016 > 1$.
A pure state $|\widetilde{T_C}\rangle$ undergoing noiseless tomography produced an expected result that is non-physical.
Unfortunately, the existence of $|\widetilde{T_C}\rangle$ implies the existence of more problematic cases, where the error in the estimated state is small enough to go unnoticed but not small enough to be negligible, especially in the context of a high state fidelity. 
In this work, we bypass this problem by using kickback tomography. Note that in principle, this effect may be mitigated by Pauli twirling the noise throughout the circuit, turning the pure state into a mixed state. We do not take that approach here -- Pauli twirling must be approximated, and reporting a fidelity would require a robust understanding of the coherent noise and Pauli twirling approximation requirements.  

\section{\label{sec:simulations-lfn}Effects of coherent noise}

\subsection{\label{sec:lfn_sims}Logical state tomography in the presence of low frequency noise - simulations}
When starting this project, we observed a non-physical norm in experiments when measuring cultivated magic states with logical state tomography. We attribute this to low-frequency phase noise in the circuit. Here, we approximate coherent phase noise in state vector simulations and show that indeed, non-physical norms are possible in this context. 
It is not surprising that coherent noise is playing a role in our experiments. Sources of such fixed detunings on the qubits could come from slow qubit frequency drift, residual magnetic fields from our control lines, or Stark shifts on the qubits through microwave crosstalk.
Our cultivation circuit is especially sensitive to such noise sources, because for a significant fraction of the time, the data qubits are idling in an entangled state, giving ample opportunity for coherent phase accumulation.
Intuitively, during the circuit, the stabilizers may be more sensitive to phase noise for one stabilizer quadrature (e.g. X) compared to the other (e.g. Z). Subsequent quantum error correction or post selection means that the final state of the logical qubit may be different depending on the tomography axis being measured, leading to inconsistent results and an apparent non-physical norm of the logical state when that noise is coherent. While Pauli twirling of the noise should render the tomography results physical (norm less than one), in practice twirling must be approximated. Because we want to rigorously measure high state fidelities, we did not trust our approximations of Pauli twirling for noise sources that are not yet fully characterized or understood, and instead developed kickback tomography for a single, self-consistent measurement.  

For simulations, we first choose a random ``detuning" for each qubit, drawn from a uniform distribution from $-200$~kHz to $200$~kHz. Then, phases are added to each qubit at every gate layer, given by the gate layer duration (35 ns, for both single qubit and two qubit gate layers) and its fixed detuning. (Note that no phases are added to data qubits that idle during measure qubit readout and reset, as we assume that the dynamical decoupling we perform on the data qubits during that time nulls the phase accumulation; see Ref.~\cite{acharya_quantum_2025} for details). This approximates circuit errors to first order in the phase noise, neglecting effects like pulse errors or gate distortions that could lead to bit flip errors or leakage. The detuning on each qubit is fixed for every run of our simulation, to approximate quasi-static noise (in future work, a more accurate simulation could change the detuning according to a $1/f$ spectrum, for example).

\begin{figure*}
    \centering
    \includegraphics[width=\textwidth]{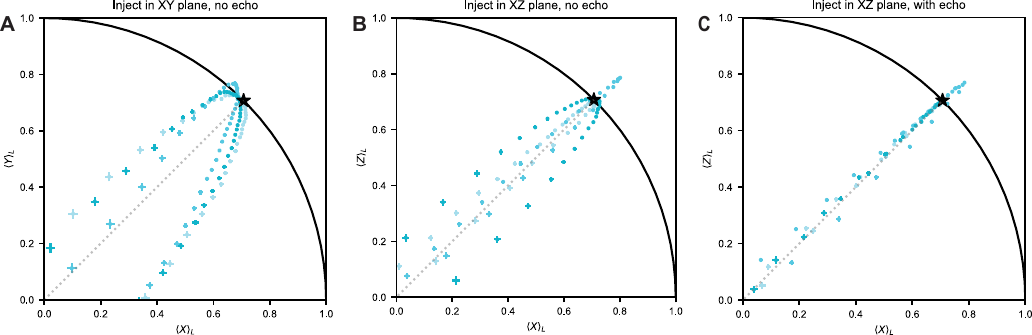}
    \caption{State vector simulation results for circuits containing state injection, cultivation, and tomography (see Fig 2(a) of the main text). The simulations include an SI1000 depolarizing noise model ($p = 3.0 \times 10^{-3}$) as well as coherent phase errors applied to each qubit at every circuit layer. The coherent errors are given by a random, fixed detuning applied to each qubit, to simulate low-frequency phase noise. Detunings are drawn from a uniform random distribution from $-200$ to $200$ kHz. We perform each simulation three times (color intensity), with three different sets of detunings, drawn from the uniform random distribution, that are consistent across panels A-C. (a) We inject in the $XY$ plane and measure the $X$ and $Y$ stabilizers during tomography to reconstruct the state. No dynamical decoupling strategy is applied. (b) We inject in the $XZ$ plane, using our injection circuit described in Sec.~\ref{sec:explicit_circuits}, and do not include any dynamical decoupling strategy. (c) Circuit used in the main text. Injection in the $XZ$ plane, with echoing added to the circuit. We observe the least amount of distortion this way, but non-physical norms are still possible, as shown.}
    \label{fig:lfn_sims}
\end{figure*}

We run simulations of our cultivation circuit, again sweeping $\theta$ as in the cyan curve of Fig.~2(b) in the main text. We include an SI1000 Pauli noise model in addition to the coherent noise, with a noise strength $p = 3.0 \times 10^{-3}$ as is done for the simulations corresponding to the gray points. In Fig.~\ref{fig:lfn_sims}, we show simulation results for three sets of random qubit detunings and for three circuit strategies. 

First, as shown in Fig.~\ref{fig:lfn_sims}(a), we inject in the XY plane instead of XZ plane, cultivate the $e^{-i\hat{\sigma}_z \pi/8} \ket{+}$ state, and measure the X and Y stabilizers during tomography to reconstruct the logical state. Because of the different sensitivities of the X and Y stabilizers to coherent phase noise, the resulting state is asymmetric across the $X = Y$ line. Additionally, non-physical norms are possible when injecting close to $\phi = \pi/4$. We attribute both effects to the coherent phases added to the circuit.

In Fig.~\ref{fig:lfn_sims}(b), we instead inject in the $XZ$ plane, using our injection circuit described in Sec.~\ref{sec:explicit_circuits}. While the logical qubit collapses closer to the $X=Z$ line, there are still deviations from it and a nonphysical norm is possible, both of which we again attribute to the coherent errors added to the circuit. As shown in the main text, Fig.~2(b), gray curve, both of these effects disappear when only simulating Pauli noise.

In Fig.~\ref{fig:lfn_sims}(c), we again inject in the $XZ$ plane our injection circuit described in Sec.~\ref{sec:explicit_circuits}. However, we also apply Pauli $\hat{X}$ gates halfway between pairs of Hadamard gates on each qubit, to ``echo" low frequency noise wherever possible. We adjust the circuit before each non-Clifford ($\hat{T}$) layer to invert these echo gates and restore the Pauli frame, preserving the desired evolution. We observe that the state is collapsed to the $X=Z$ line, implying that the phase noise has largely been mitigated. However, for some sets of detunings, non-physical norms can still be observed, as shown. This is why we do not quote a state fidelity using logical state tomography, and develop and implement kickback tomography instead.
\subsection{\label{sec:with_without_echos} Magic state performance with and without echos - experimental data}
\begin{figure}
    \centering
    \includegraphics[width=0.3\linewidth]{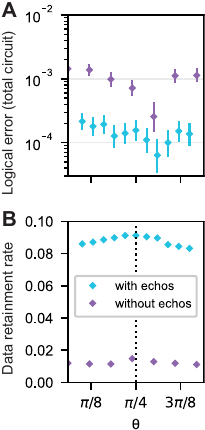}
    \caption{Logical error probability (full circuit) as measured via kickback tomography, and data retainment rates, with and without echos to filter coherent phase noise.}
    \label{fig:with_without_echo}
\end{figure}

As discussed in Sec.~\ref{sec:lfn_sims}, we add echos to the circuit by finding consecutive pairs of Hadamards on each qubit and applying a $\pi$ pulse halfway between them. We study the difference in performance as measured in kickback tomography in Fig.~\ref{fig:with_without_echo}. In both cases, RL calibration is performed on each circuit for about four hours. The results suggest that without echos, our performance is dominated by low-frequency phase noise.

To determine how much our state performance may be limited by coherent phase noise even with echoing, we performed additional simulations of kickback tomography, using the same random distribution for qubit detunings as in Sec.~\ref{sec:lfn_sims}, i.e., uniform random distributed from -200 to 200 kHz (similar to our expected low frequency noise magnitude), and without any depolarizing noise. We found an upper bound of $\ket{T}$ state infidelity of order $\sim 10^{-5}$, well below our experimental results. This suggests that our magic state performance when using echoing is limited by gate error and decoherence, not quasi-static detunings.
\section{\label{sec:simulations}State vector simulation methods}

In this section we discuss the numerical techniques used to simulate the magic state cultivation, kickback tomography, and grafting experiments. These experiments are more challenging to simulate than quantum memory experiments because they involve $\hat{T}$ gates. $\hat{T}$ gate unitaries are not elements of the Clifford group and are therefore not compatible with standard Clifford simulation methods~\cite{Aaronson-Gottesman-2004}. Instead, they require a “fully quantum” description of the system’s state vector as it propagates throughout the experimental circuit. For the distance-3 cultivation and kickback tomography experiments (Figs. 2 and 3), we used the \texttt{kraus-sim} quantum trajectories simulator. While the experiments used 12(14) measure, helper, and data qubits for Fig. 2(3), to speed up the simulation we implemented a modified Kraus operator reordering algorithm to ensure that, averaging over the circuit layers, 8.4(10.2) qubits were required in memory\cite{obrien_DensitymatrixSimulationSmall_2017a, marshall_IncoherentApproximationLeakage_2023}. This is described in Subsection~\ref{sec:reordering}. 

To simulate the distance-5 (54 qubit) grafting experiment, we used a different simulation technique. This technique was implemented in the simulator \texttt{qevol}, an extension of the Pauli+ simulator (described in ref.~\cite{google2023suppressing}) that maintains superpositions of stabilizer states. It is particularly amenable to simulations of these experiments in this work given the modest number of $\hat{T}$ gates used. The simulator implementation is described in detail in Subsection~\ref{sec:lfn}.

\subsection{\texttt{kraus-sim}}
\label{sec:reordering}
Here we discuss a numerical method to optimize the qubit-count overhead in a (noisy) circuit simulation, which is of particular relevance to quantum trajectories simulations which store the full state vector. The aim of the optimization is, given a particular circuit description as a sequence of quantum channels, to find an equivalent ordering of the operations, such that the number of qubits required to be kept in memory is minimized. Two orderings are equivalent if one can be transformed into the other through a sequence of swaps of consecutive commuting operations. In practice, we only swap operations which trivially commute, acting on disjoint qubits. 
This takes advantage of the observation that after measuring a qubit, it can be removed from the simulation, and only reintroduced later when it's required. 
This is directly applicable to error correction simulations, where measure qubits are repeatedly measured and reset, allowing them to be periodically removed from memory. In fact, it is known that in the repetition and surface codes, it is possible to perform a simulation with at most one measure qubit in memory at any one time \cite{obrien_DensitymatrixSimulationSmall_2017a, marshall_IncoherentApproximationLeakage_2023}. In particular, for the rotated surface code this reduces the active qubit-count from $2d^2-1$ to $d^2+1$.
Though in general finding an optimal solution to this task appears to be a non-trivial problem, we can in practice make good progress using a heuristic optimization method.

In order to facilitate the removal and addition of qubit degrees of freedom from a circuit, we convert all measurement operations to \textit{destructive measurements} followed by \textit{creative resets}, which reduce or increase the qubit count by one, respectively. These are formally defined in \cite{marshall_IncoherentApproximationLeakage_2023}, but can simply be thought of as operations with non-square Kraus operators, such as $\{\langle 0|, \langle 1|\}$ for a destructive measurement. Note that reset to either $|0\rangle$ or $|1\rangle$ can be conditioned on a prior measurement. However, in the setting of the relevant error correction circuits, a reset deterministically prepares the state $|0\rangle$.

\begin{figure}
    \centering
    \includegraphics[width=0.5\linewidth]{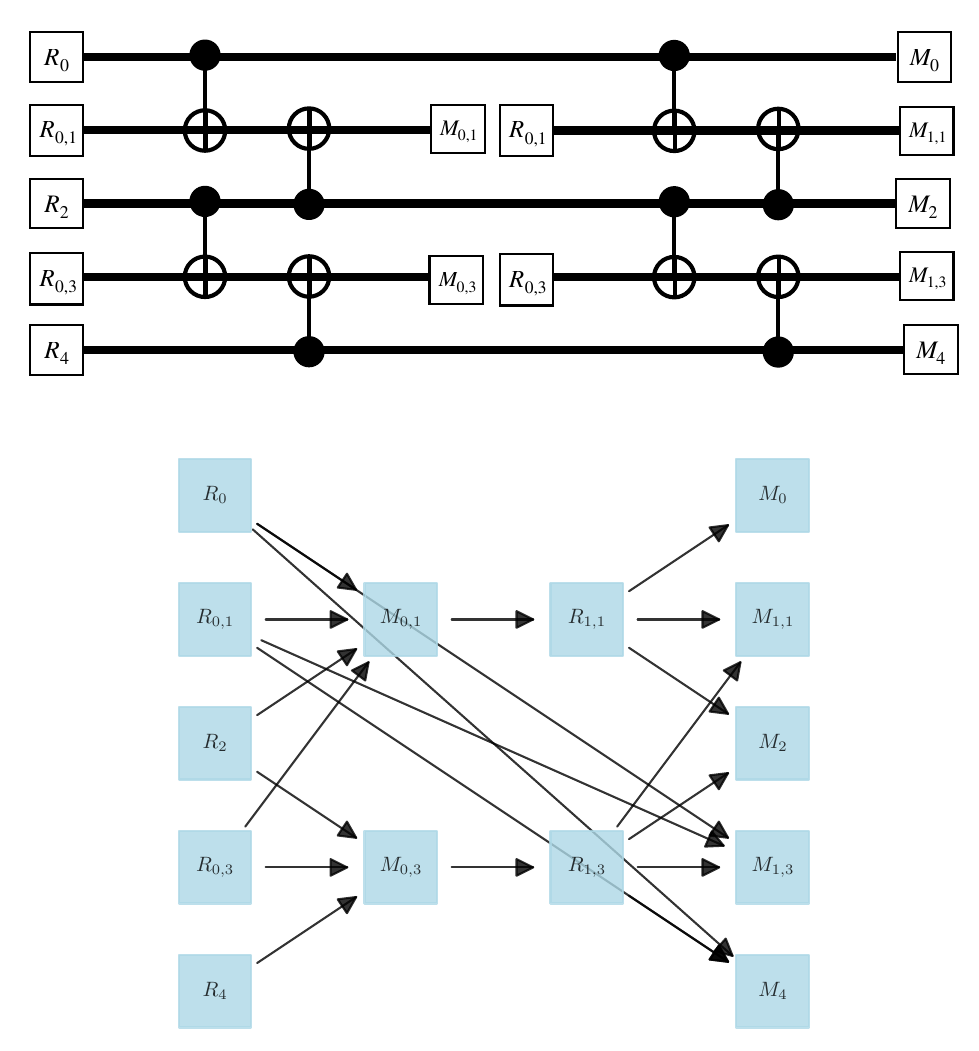}
    \caption{(Top) Two QEC cycles of distance 3 repetition code circuit. (Bottom) Simplified DAG of only the reset/measurement operations.}
    \label{fig:circuit-dag}
\end{figure}

We define a noisy quantum circuit as a sequence of quantum channels $(\mathcal{E}_1, \mathcal{E}_2, \dots, \mathcal{E}_N)$, which are applied from left to right. This sequence is then decomposed into a directed acyclic graph (DAG) as follows: Each node $j$ of the graph is associated with channel $\mathcal{E}_j$. A directed edge $j \rightarrow k$ is assigned whenever, for $j<k$:
\begin{itemize}
    \item There exists a qubit $q$ acted on by both $\mathcal{E}_j$, $\mathcal{E}_k$
    \item There is no $l$ such that $j < l < k$, where $\mathcal{E}_l$ acts on $q$.
\end{itemize}
In particular, there is some qubit $q$, such that operation $j$ will first act on it, followed directly by operation $k$. We call a node $j$ `upstream' of $k$, if there is a path from $j$ to $k$ in the DAG. 
We can observe that \textit{any} ordering of the operations is physically equivalent, so long as the ordering implied by the above edges is conserved. In particular, this means we can freely add edges to the DAG (so long as it indeed remains a DAG) without breaking any causality constraints. Going the other way, a sequence of operations consistent with the DAG can be found by performing a topological sort.
We show an example of a circuit and a DAG in Fig.~\ref{fig:circuit-dag}. In the DAG, we have pruned the operations that are not measurements or resets for simplicity, but keep track of the implied causality (for the purposes of reordering, one only requires the DAG of resets/measurements).

Now, notice that one could solve the problem in question by brute force; given a valid ordering of measurements (the measurement order is consistent with the circuit DAG), it is simple to figure out the maximal number of qubits in memory at any one time. In particular, the first measurement in the ordering uniquely determines the requisite upstream reset operations to be included beforehand (in a valid ordering, the first measurement has no upstream measurements). After the measurement, one can remove the associated degree of freedom (qubit), and consider the second measurement, which requires including additional reset operations. One can go through the list of measurements, and construct the full circuit order, and thus count the number of qubits in memory at any one time. Doing this for all valid measurement orderings would result in the optimal circuit. Of course, this is infeasible, since for $m$ measurements, there are $m!$ measurement orders. Whilst many of these will not be valid orderings, the number of valid combinations is in most cases likely still prohibitive.

Instead, we consider a heuristic approach, where measurements are added greedily into the circuit. To begin with, we start with an empty list $\mathcal{C}$, to be populated by the order of resets and measurements. We additionally keep track of $\mathcal{M}$, as the set of measurements that are not yet included in $\mathcal{C}$ (initially $\mathcal{M}$ consists of all measurements).
Then we repeatedly apply the following steps until all measurements have been placed:
\begin{enumerate}
    \item Generate the subset $\mathcal{M}'\subseteq \mathcal{M}$, such that $M\in \mathcal{M}'$ has no upstream measurements in $\mathcal{M}$. That is, $M\in \mathcal{M}'$ has no upstream measurements that still need to be included in the circuit.
    \item Pick the element $M_{\mathrm{opt}}$ in $\mathcal{M}'$ that introduces the fewest number of resets to the circuit, as determined by the number of upstream resets of $M_{\mathrm{opt}}$ that are not already in $\mathcal{C}$. Pick randomly in case of a tie.  Append all (new) upstream reset operations of $M_{\mathrm{opt}}$ to $\mathcal{C}$, followed by $M_{\mathrm{opt}}$ itself. Remove $M_{\mathrm{opt}}$ from $\mathcal{M}$.
\end{enumerate}
Due to the randomness of tie-breaking, it is often advantageous to run the above routine several times. To enforce the constraints imposed by the new ordering $\mathcal{C}$, the relevant edges implied by $\mathcal{C}$ are added to the circuit DAG. Performing a topological sort will result in the optimized circuit.

\begin{figure}
    \centering
    \includegraphics[width=\linewidth]{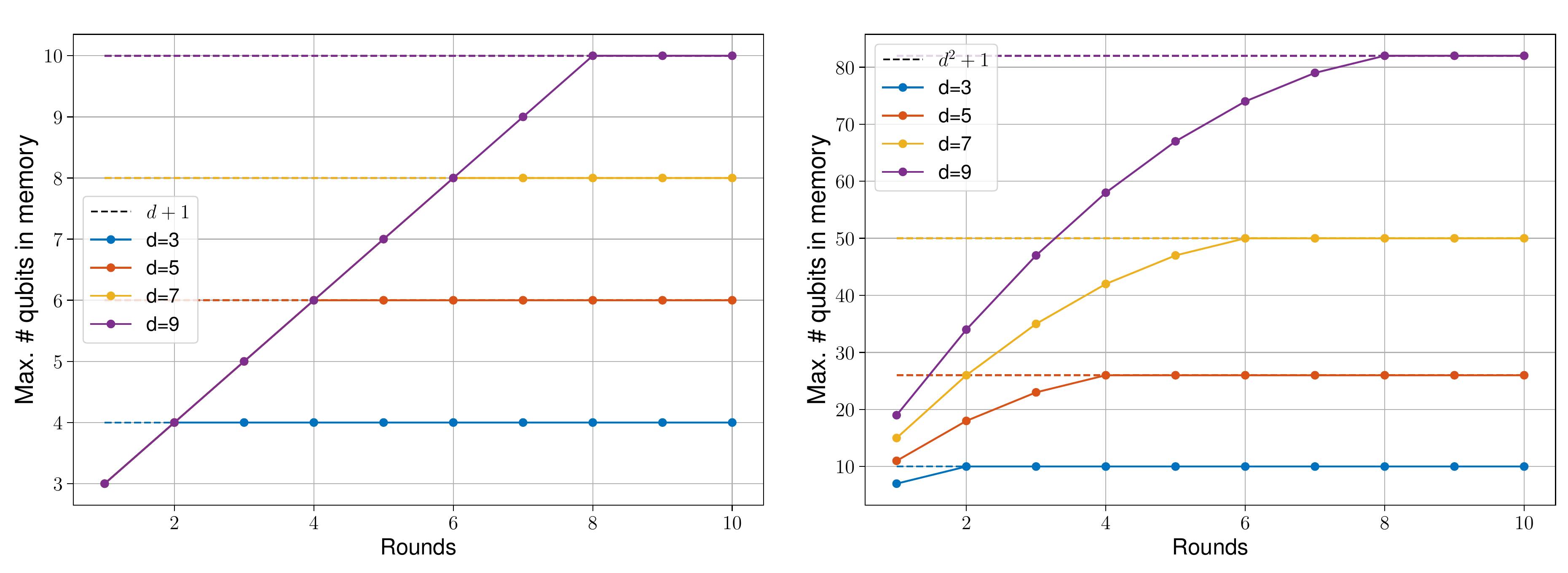}
    \caption{(Left) Optimized qubit count for repetition codes of varying code distances $d$, as a function of circuit depth (number of QEC cycles). (Right) Rotated surface code. We see in both cases, for deep enough circuits ($\ge d-1$ cycles), the result tends to a single measure qubit, namely $d+1$ and $d^2+1$ respectively, but for short circuits, the requirements can be significantly less; a 1 cycles surface code only requires $2d+1$ qubits in memory.}
    \label{fig:reordered-qubit-count}
\end{figure}

When applied to Fig.~\ref{fig:circuit-dag}, initially we have $\mathcal{C}=[]$ and $\mathcal{M}'=[M_{0,1}, M_{0,3}]$. Note, the remaining 5 measurements are not included in $\mathcal{M}'$, since they all require at least one of $M_{0,1}, M_{0,3}$ to be included first, as they are upstream. The first measurement added to $\mathcal{C}$ will be $M_{0,3}$, since it has only 3 upstream resets, in comparison to $M_{0,1}$ which has 4. After this initial step, we have $\mathcal{C}=[R_2, R_{0,3}, R_4, M_{0,3}]$, $\mathcal{M}'=[M_{0,1}, M_{1,3}, M_4]$. As before, the remaining measurements ($M_0, M_{1,1}, M_2$) are not included in $\mathcal{M}'$, as they all have $M_{0,1}$ upstream. Next, $M_{0,1}$ would be added, which requires two new qubits (from $R_{0}, R_{0,1}$). In contrast, $M_{1,3}$ and $M_4$ would require three new qubits ($R_0, R_{0,1}, R_{1,3}$). At this point, $\mathcal{M}'$ consists of the final 5 measurements, as all of their upstream measurements have already been included in the circuit.
Continuing this process, we arrive at the full order of resets/measurements (ties were broken arbitrarily):
$$\mathcal{C} = [R_2, R_{0,3}, R_4, M_{0,3}, R_0, R_{0,1}, M_{0,1}, R_{1,3}, M_{1,3}, M_4, R_{1,1}, M_0, M_{1,1}, M_2].$$
This requires at most 4 qubits to be kept in memory (directly after $R_{0,1}$).
In order to arrive at the full circuit order, one adds the above directed edges to the circuit DAG, to enforce the ordering ($R_2 \rightarrow R_{0,3} \rightarrow R_4 \rightarrow M_{0,3}\rightarrow \dots$).
Performing a topological sort of the new DAG will yield the qubit-count optimized circuit.
We show results in Fig.~\ref{fig:reordered-qubit-count} for repetition and surface codes, which demonstrates significant savings for shallow circuits.


\subsection{Restricted-rank quantum trajectory simulator}
\label{sec:lfn}

For faster simulations, we have also used a restricted-rank quantum
trajectory simulator called {\tt qevol}.  Unlike {\tt kraus\_sim}
which represents an $n$-qubit state vector $\Psi$ in terms of $2^n$
complex coefficients, {\tt qevol} maintains an expansion in the basis
of a stabilizer group that is in some sense optimal for the given
state vector $\Psi$.  At any time (and this is different from the
stabilizer-state decomposition in
Refs.~\onlinecite{Bravyi-Gosset-2016,Bravyi-etal-sim-2019}), there is
only one stabilizer group with $n$ generators, so that a $k$-qubit
Clifford unitary (which is implemented by transforming the basis) has
complexity of order ${\cal O}(kn)$, independent of the number of
coefficients in the decomposition.  The same applies to an arbitrary
$k$-qubit Pauli channel.

Other operations, like single-qubit measurements and reset
operations, or general few-qubit channels, require that coefficients
in the expansion be modified.  To make the change as local as
possible, a set of tableau generators is chosen so that only a few are
supported on the qubits in question.  This allows to define an
auxiliary local representation of the original state which involves
only a few qubits, and the channel be applied locally.  After that, a
new local tableau is chosen, for a near-optimal description of the
modified state vector, and the global tableau operators are updated.
As explained in more detail below, for an initial state with $m$
non-zero coefficients and a channel acting on $k$ qubits, the
complexity is ${\cal O}(kN^2)$ to update the tableau, and up to
${\cal O}(4^{2k})+{\cal O}(N 2^{2k})$ for the local-basis operations,
where $N=\max(n,m)$, and some additional logarithmic factors have
been omitted.

\subsubsection{State representation used by {\tt qevol}}
\label{sec:state}
Internally, {\tt qevol} maintains a Gottesman
tableau\cite{Aaronson-Gottesman-2004}, two sets of $n$ commuting
Hermitian $n$-qubit Pauli operators, stabilizer generators $G_i$ and
destabilizers $D_j$, $1\le i,j\le n$, such that $G_i$ and $D_j$
anticommute if and only if $i=j$.  Explicitly, the state decomposition
has the form,
\begin{equation}
  \Psi= \sum_s C_s D^s\ket{\cal S}\equiv \sum_s C_s \prod_{i=1}^n
  D_i^{s_i}\ket{\cal S},\label{eq:Psi}
\end{equation}
where the stabilizer state $\ket{\cal S}$ (defined up to an overall
phase) is a joint $+1$ eigenstate of the stabilizer generators $G_i$
(and the elements of the stabilizer group
${\cal S}\equiv \langle G_1,G_2,\ldots, G_n\rangle$), and complex
coefficients $C_s$ are associated with the binary syndrome vectors $s$
with components $s_i\in\{0,1\}$, $1\le i\le n$.  Generally, such an
expansion contains $2^n$ coefficients; in practice the number $m$ of
non-zero coefficients can be much smaller.

It is easy to verify that for any $n$-qubit Clifford unitary $U$, the
state $U\Psi$, up to an overall phase, can be written in the form
Eq.~(\ref{eq:Psi}) with the same coefficients $C_s$, but transformed
generators, $D_i\to UD_iU^\dagger$, $G_i\to UG_iU^\dagger$.  In
particular, when $U$ is a Pauli operator, such a transformation
reduces to a change of some signs.  This
implies that a $k$-qubit Clifford
unitary or a Pauli channel can be implemented efficiently, with the
complexity ${\cal O}(kn)$~\cite{Aaronson-Gottesman-2004}.

A representation (Eq.~\ref{eq:Psi}) is not unique.  Canonical operations
modifying the set of generators but preserving commutation relations,
the stabilizer group, and the state vector include:
\begin{itemize}
\item {\tt swap}$(i,j)$: swap generators and syndrome bits
  $(G_i,D_i,s_i)$ and $(G_j,D_j,s_j)$;
\item {\tt GG}$(i,j)$: $G_i\to G_i G_j$, $D_j\to D_j D_i$,
  $s_i\to s_i+ s_j$;
\item{\tt DG1}$(i)$: $D_i\to -iD_i G_i$, and also $C_s\to iC_s$ iff
  $s_i=1$;
\item{\tt DG2}$(i,j)$: $D_i\to D_i G_j$, $D_j \to D_j G_i$, and also
  $C_s \to -C_s$ iff $s_i=s_j=1$.
\end{itemize}
These operations, each with complexity ${\cal O}(N)$, preserve the
magnitudes of the coefficients.

\subsubsection{Canonical local entanglement form}
Application of a quantum channel, a completely positive
trace-preserving map acting on $k$ qubits in a set
${\cal A}\subset \{1,2,\ldots,n\}$, starts with transforming the
tableau to a canonical local entanglement (CLE)  form. To simplify
the presentation, take ${\cal A}=\{1,2,\ldots,k\}$.  Then, a CLE form is
defined as follows:
\begin{itemize}
\item The first $k$ destabilizers of weight $1$,
  $D_i=P_i\in \{\pm X_i,\pm Y_i,\pm Z_i\}$, are supported on the
  corresponding qubits, $\mathop{\rm supp} D_i=\{i\}$, $i\in {\cal A}$, while the
  remaining destabilizers have no support on ${\cal A}$,
  $\mathop{\rm supp} D_j \cap {\cal A}=\emptyset$.
\item Exactly $k+p$ stabilizer generators should have any support in
  ${\cal A}$, where $0\le p\le k$ is the entropy of bipartite
  entanglement\cite{Fattal-Cubitt-Yamamoto-Bravyi-Chuang-2004} of the
  region ${\cal A}$ with its complement in the stabilizer state
  $\ket{\cal S}$.  Further, the first $k$ stabilizer generators should
  commute with each other when punctured to ${\cal A}$,
  $G_i[{\cal A}]G_j[{\cal A}]=G_j[{\cal A}]G_i[{\cal A}]$,
  $ i,j\in {\cal A}$, while each of the remaining $p$ generators
  should anticommute with exactly one of the first $k$,
  $G_{i(j)}[{\cal A}]G_{j+k}[{\cal A}]=-G_{j+k}[{\cal
    A}]G_{i(j)}[{\cal A}]$, where $1\le j\le p$ and the pivot index
  $i(j)$ is a strictly increasing function of its argument $j$,
  $i: \{1,\ldots, p\}\to {\cal A}$.
\end{itemize}
Here, $G_j[{\cal A}]$ is a $k$-qubit Pauli operator obtained from $G_j$
by dropping all single-qubit factors outside ${\cal A}$.  Since each
of $D_i=P_i$, $i\le k$, anticommutes only with the corresponding
$G_i$, these commutation relations imply that single-qubit Pauli
operators along the diagonal are $G_i[i]= Q_i\not\in\{I,P_i\}$,
$i\le k$, off the diagonal $G_i[j]\in\{I,P_i\}$, $j\neq i\le k$, while each of the
operators $G_{k+j}[{\cal A}]$, $j\le p$, contains only one non-trivial
term, $G_{k+i(j)}[{\cal A}]=\pm P_j$ for $j\le p$.  That is, the
operators $G_j[{\cal A}]$, $j\le k$, generators of the local
stabilizer group, are in a graph state form~\cite{Hein-2006}, up to a
local Clifford transformation, while the operators
$G_{j+k}[{\cal A}]$, $j\le p$, coincide with the destabilizers
$D_{i(j)}$, up to a sign.

Reduction to a CLE form is done with the help of canonical operations
discussed in Sec.~\ref{sec:state}; it is similar to the Gauss
elimination algorithm.  The corresponding complexity is
${\cal O}(N^2)$ per column, or ${\cal O}(kN^2)$ for a set of $k$
qubits.

\subsubsection{Local tableau construction}
\label{sec:local-tab}

Given the constructed CLE form of a tableau with $k+p$ non-trivial
rows $G_i[{\cal A}]$, we add $p$ ancillary qubits and multiply the
punctured generators $G_i[{\cal A}]$ by the corresponding $X$ and $Z$
operators to ensure commutativity.  The operators $G_i[{\cal B}]$,
where set ${\cal B}$ is the complement of ${\cal A}$ are also
modified, to simplify bookkeeping:
\begin{enumerate}
\item Multiply every pivot row among the first $k$ by a $Z$ ancillary
  operator, to get local generators:
  $G_{i(j)}[{\cal A}] \overline{Z}_j$, $j\le p$, store
  $G_{i(j)}[{\cal B}] \overline{Z}_j$.  The remaining (non-pivot) rows
  are entirely
  supported on
  ${\cal A}$ and require no modification~\cite{Fattal-Cubitt-Yamamoto-Bravyi-Chuang-2004}.
\item Multiply each of the $p$ additional rows by an $X$ ancillary
  operator, $G_{j+k}[{\cal A}] \overline{X}_j$, store
  $G_{j+k}[{\cal B}] \overline{X}_j$, $j\le p$.
\item Add destabilizer rows $D_{j+k}=\overline{Z}_j$ formed by
  ancillary qubits only, $j\le p$.
\end{enumerate}
With this representation, a local decomposition of the state is
readily constructed, by splitting the syndrome vectors $s\to s'|s''$,
where $s'$ has only the syndrome bits corresponding to the first $k+p$
stabilizer generators with a support on the qubits in ${\cal A}$.  The
remaining syndrome bits are not affected by the transformation.  An
example of such a decomposition is shown in
Fig.~\ref{fig:clef-decomposition}.

\begin{figure}[htbp]
  \centering
    \begin{minipage}[c]{3in}
\begin{verbatim}
   1234       12a    a34
G1 ZZZZ       ZZZ    ZZZ
G2 XX..       XX.     ..
G3 X.X.       X.X    XX.
G4 ..XX       ...     XX
   ---- ----> --      -- 
D1 X...       X..    X..
D2 .Z..       .Z.    ...
D3 ..ZZ       ..Z    .ZZ
D4 ...Z       ...    ..Z
\end{verbatim}
  \end{minipage}
  \caption{Example of a local tableau on qubits ${\cal A}=\{1,2\}$ and
    an ancillary qubit ${\tt a}$ from a 4-qubit tableau in a CLE form.
    Top row shows qubit indices, rows labeled {\tt G}$i$ and {\tt
      D}$i$ with $1\le i\le 4$, respectively, correspond to stabilizer
    generators and destabilizers.}  
\label{fig:clef-decomposition}
\end{figure}
After constructing a local tableau on $k+p$ qubits, any state vector
in the form (\ref{eq:Psi}) can be written as a density matrix $\rho$
on $k+p$ qubits, namely, the qubits in ${\cal A}$ and the $p$
ancillary qubits added.  The construction uses only the first $k+p$
syndrome bits; the remaining syndrome bits are summed over.  To
apply an arbitrary two-qubit unitary operation,
$\rho\to (U\times I)\rho(U^\dagger\times I)$, we
\begin{enumerate*}[label=(\alph*)]
\item construct a unitary matrix transforming from the basis specified
  by the local tableau to the $Z$-basis on $k+p$ qubits,
\item transform the
basis vectors using only the first $k+p$ bits of the syndrome, 
\item apply the unitary $U\times I_p$, 
\item select new local stabilizer
  basis, 
\item construct yet another unitary transforming from the $Z$-basis
  to the modified local tableau generators chosen, 
\item transform the coefficients, dropping those whose magnitude is
  below certain cut-off $\varepsilon$, typically
  $\varepsilon=10^{-4}$, and, finally,
\item update the global tableau by tracing out the ancillary qubits.
\end{enumerate*}
\subsection{Trimming the coefficient list}
The number of non-zero coefficients may increase after a unitary or a
CPTP map is applied.  If the number of non-zero coefficients exceeds
the desired maximum, the coefficient list is trimmed, by removing a
sufficient number of the coefficients smallest in magnitude.  In the
present simulations this was not necessary, as the number of terms in
the expansion (\ref{eq:Psi}) remained well below the set limit,
$m_{\rm max}=10^3$.

\section{Validating kickback tomography}

In this section, we report additional discussion and simulations that validate kickback tomography (KT) as a metrological tool for measuring the lower bound on $\ket{T}$ state fidelities. In KT, the QEC cycles are crucial for detecting and post selecting against errors. Note that one could instead decode in a trade-off of performance for speed, beyond the scope of this work. In particular, the stabilizer measurements ensure that the noise can be described with a Pauli model, digitizing it as usual when running a QEC memory experiment. For example, suppose there was an under or over-rotation of the $\ket{T}$ gate in the first gate of the $\hat{H} = \hat{T}^\dagger \hat{X} \hat{T} $ implementation. Because cultivation comprises an implementation of the \textit{fault-tolerant}, logical gate $\hat{H}_L$, this under or over-rotation would appear as a physical qubit gate error, which would then be caught by the stabilizer rounds. Chains of errors that are evaded by the stabilizers or cultivation detectors result in the logical error probability, which we report. 
\begin{figure}
    \centering
\includegraphics[width=0.6\linewidth]{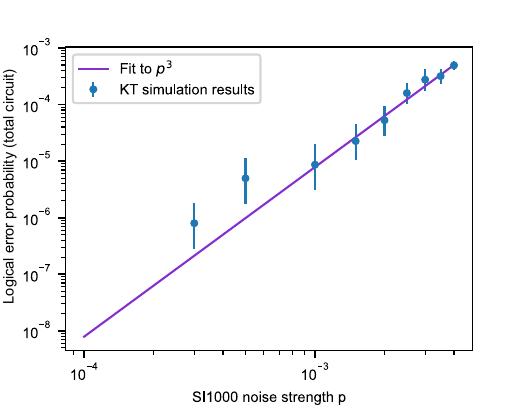}
    \caption{Validating that the upper bound of the $\ket{T}$ state we measure with kickback tomography scales as the cube of the noise strength. We simulate the circuit in Fig. 3 (cyan curve) of the main text, measuring a cultivated $\ket{T}$ state with KT. We sweep the noise strength $p$ of our SI1000 simulations and show that the extracted logical error probability scales as $p^3$. This is expected, because cultivation increases the fault distance of the state to three. The error bars represent the 68\% confidence interval using a Clopper Pearson statistical test.}
    \label{fig:validating_KT}
\end{figure}

We also confirm that the upper bound in the $\ket{T}$ state error that we measure with KT scales as the cube of the noise strength, as expected since cultivation improves the fault distance of the magic state to three \cite{gidney_magic_2024}. We run simulations using {\tt kraus\_sim}, with an SI1000 depolarizing noise model \cite{gidney_fault-tolerant_2021}, sweeping the noise strength $p$. The simulated circuit is same one as described in Fig. 3 of the main text with cultivation (cyan curve). We inject an angle $\theta = \pi/4$ to specifically study the best-case $\ket{T}$ state error (and as would be done in practice when running quantum algorithms). Fig.~\ref{fig:validating_KT} shows that the extracted logical error probability is described by a fit proportional to $p^3$. This validates that our KT metrology tool is estimating the $\ket{T}$ state fidelity as expected.   
\section{Kickback tomography with interleaved QEC cycles}
\begin{figure}
    \centering
\includegraphics[width=0.8\linewidth]{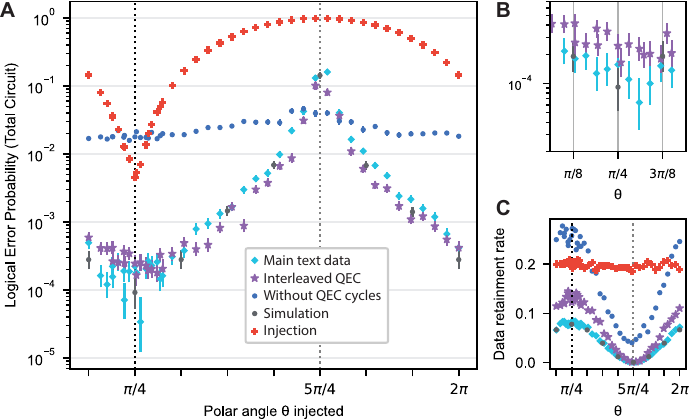}
    \caption{Adding an ``interleaved QEC" circuit to the kickback tomography results (purple stars). This circuit has a fault distance of 3, like the one in the main text, but the second and fourth QEC cycles are removed. (a) As expected, we observe similar fidelity for the interleaved QEC case as the usual case, which is because both circuits have the same fault distance. (b) Zooming in and taking more statistics, we can distinguish slightly better performance when including all QEC cycles, which we attribute to smaller detection regions in the circuit as the stabilizers are measured more frequently. (c) We observe slightly higher post selection rates for the interleaved qec circuit case, because there are few detectors to post select on.}
    \label{fig:interleaved_qec_big}
\end{figure}

\begin{figure}
    \centering
    \includegraphics[width=0.4\linewidth]{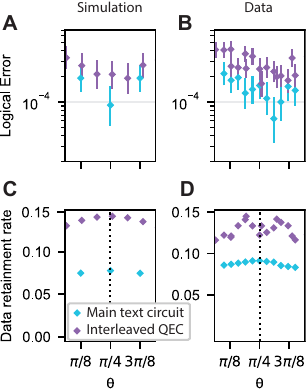}
    \caption{Comparing experiment and simulations for kickback tomography circuits with all QEC cycles and with interleaved QEC cycles. For simulations, we use an SI1000 depolarizing noise model with noise strength $p = 2.3 \times 10^{-3}$, as in the main text. (a, b): Simulation and experimental data for logical error in kickback tomography. Cyan data in (b) is the same as in the main text as well as in Fig.~\ref{fig:interleaved_qec_big}(b). (c, d): Simulation and data for post selection rates for the two circuits. We observe qualitative agreement between simulation and experiment. We attribute the improvement when retaining all QEC cycles to be from reducing the number of detection event mechanisms that can contribute to each detector.}
    \label{fig:interleaved_qec}
\end{figure}

Instead of running kickback tomography with QEC cycles between every cultivation round, we may remove two QEC cycles and still maintain a fault distance of three. Specifically, an ``Interleaving QEC" circuit may remove the second and fourth QEC cycles from the original circuit in Fig. 3(a), while maintaining fault tolerance. We show experimental results when running this interleaved-QEC circuit in Fig.~\ref{fig:interleaved_qec_big}, with comparisons to simulations in Fig.~\ref{fig:interleaved_qec}.
We observe slightly higher error for the interleaved-QEC curve, which is reproduced in simulations, as shown in Fig.~\ref{fig:interleaved_qec}. We attribute the improved performance when including all QEC cycles to the fact that there are fewer contributing error mechanisms per detector, leading to more errors being accurately detected and post selected. Similarly and consistently, the data retainment rates are slightly higher with the interleaved case. 
\section{Memory experiment in the grafted code}
\begin{figure}
    \centering
    \includegraphics[width=0.6\linewidth]{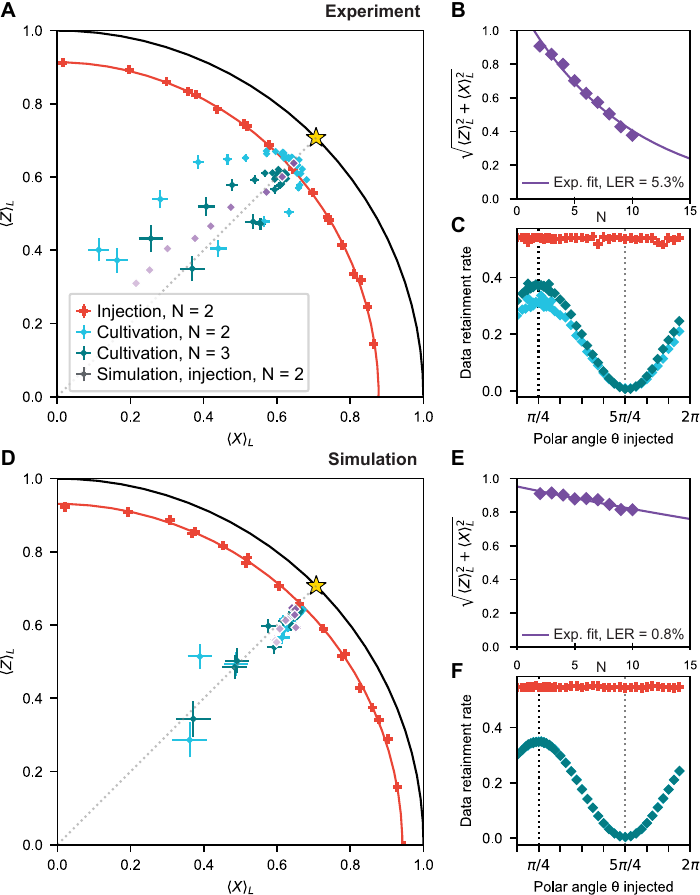}
    \caption{Memory experiment in the grafted code. (a-c): experimental data, (d-f) bottom row: simulations (SI1000 $p = 2.3 \times 10^{-3}$). We inject a variable state, cultivate it in the $d=3$ color code, then graft it to a surface code and idle for $N$ QEC cycles.  (a, d): Cultivation graft experiment as described in Fig 4 of the main text, but with increasing QEC cycles $N$ at the $\ket{T}$ state (purple diamonds, color lightness, increasing $N$). (c, f): post selection rates as a function of injected angles. Same post selection technique as described in the main text. (b, e): Logical error rate (LER) as a function of $N$. We observe worse LER than simulated, likely due to leakage as described in the text and subsequent figures.}
    \label{fig:memory_expt}
\end{figure}

\begin{figure}
    \centering
    \includegraphics[width=\linewidth]{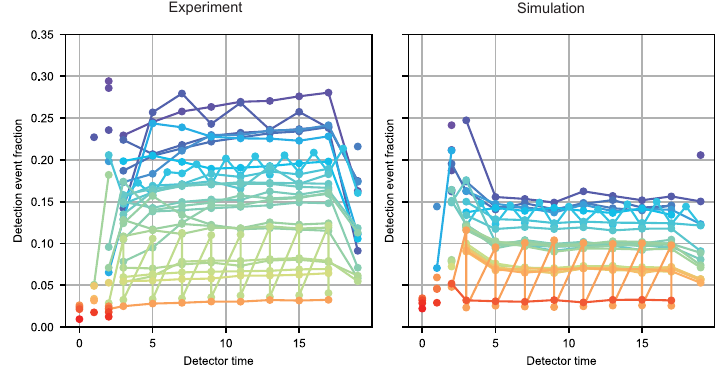}
    \caption{Detection fractions for detectors in the $N=9$ grafting circuit. We observe a slow increase in detection fractions for some detectors over time in the experimental data (left), suggestive of long-lived population excitations outside the computational subspace. This effect does not appear in simulations (right), which do not include leakage.}
    \label{fig:memory_expt_detfracs}
\end{figure}

In practice, it will be necessary to idle in the grafted code for some number of QEC cycles, either when using the magic state in gate teleportation (lattice surgery), or when escaping in the cultivation protocol \cite{gidney_magic_2024}. Although we expect to graft to much larger distances in the future, with improved logical error rate, here we run a memory experiment to start studying the cost of QEC cycles and develop our capability to do so. We inject specifically at $\theta = \pi/4$, cultivate the $\ket{T}$ state, and extend the number $N$ of QEC cycles (Fig \ref{fig:memory_expt}(a, b), purple diamonds). From those data, we extract the logical error rate (LER) for the $\ket{T}$ state in the grafted code. We also run state vector simulations of the full grafting experiment shown in Fig. 4 of the main text, also including the memory experiment (Fig \ref{fig:memory_expt}(d, e, f)). Simulations are performed using {\tt qevol}, with a simple depolarizing noise model, SI1000, $p = 2.3\times10^{-3}$, as in Fig. 4 of the main text. We observe that the LER in our experiment is about seven times higher than expected from simulations (Fig \ref{fig:memory_expt}(b, e)). It is likely that the main source of disagreement is population leakage. In particular, leakage appears as a slow increase in detection event likelihood over time within the circuit, which is seen in the experimental data of Fig. \ref{fig:memory_expt_detfracs}, but not in the simulation results (which do not include leakage approximations). 

\begin{figure}
    \centering
    \includegraphics[width=\linewidth]{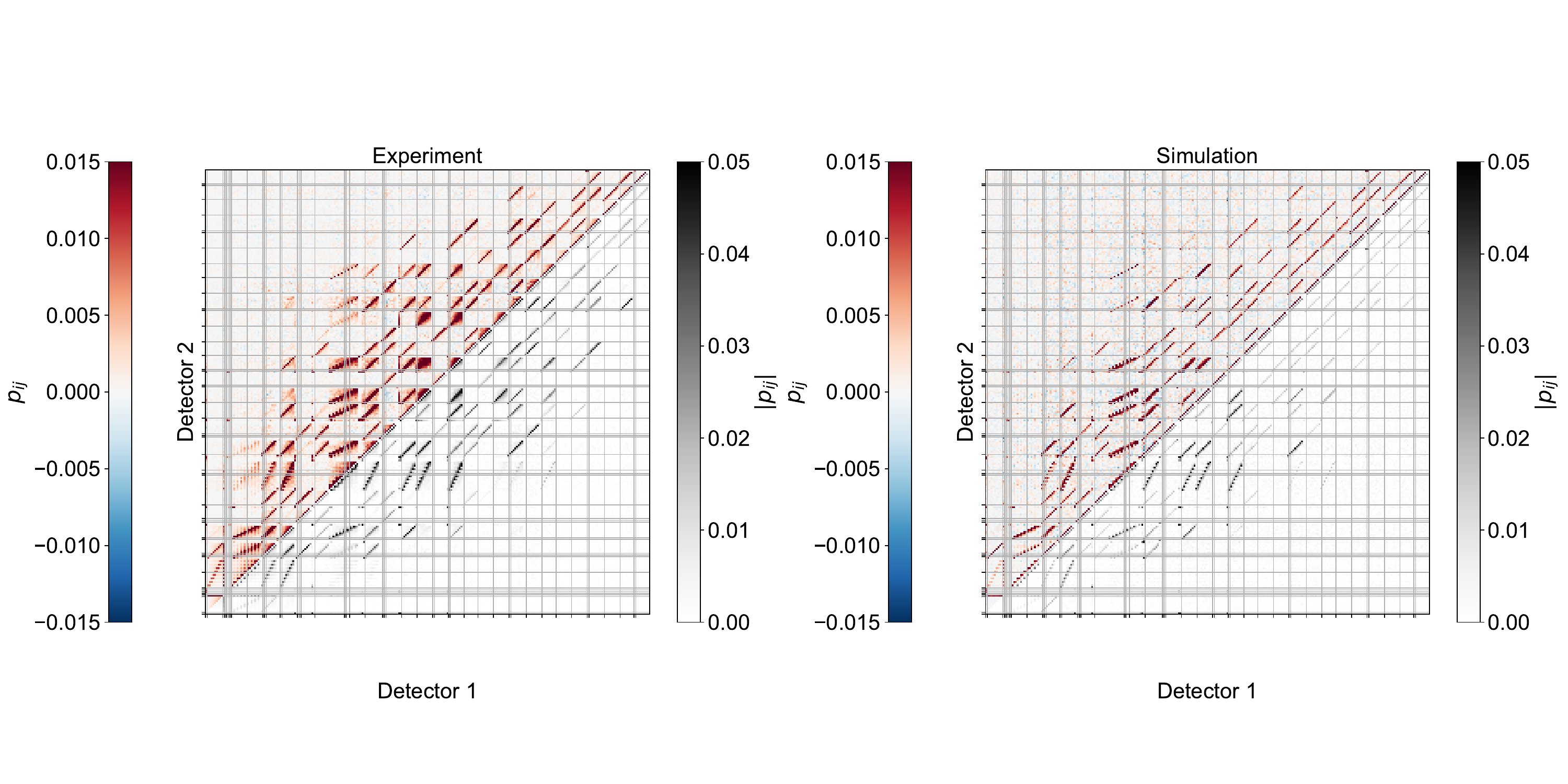}
    \caption{Correlation between detectors in a grafting memory experiment with $N = 9$ QEC cycles. Experimental data, left plot, simulation results, right plot. The detection space coordinates are on the axes, and the time coordinate is the pixel value for each cell of the 2D grid. The ``bleeding" effect observed in the data (not seen in simulation) typically corresponds to population leakage, because it is time-correlated. }
    \label{fig:memory_expt_pij}
\end{figure}

Additionally, leakage manifests as time correlations between detectors. We plot the covariance between detectors in Fig.~\ref{fig:memory_expt_pij}, and observe signatures of leakage. Specifically, the detector space coordinates are plotted on the $x$ and $y$ axes, and the time coordinates (given by the QEC cycle index within the circuit) are within each $x$ and $y$ box on the grid. Since the results are symmetric across the $x=y$ line, we plot the nominal correlation value above $x=y$ and the absolute value of it below. In the experimental data (Fig.~\ref{fig:memory_expt}, left), for some $x$ and $y$ coordinates, we observe a ``bleeding" effect, namely strong time correlations that result in the box corresponding to $x$ and $y$ appearing as a dark color. Generally, these correlations are caused by leakage, which is time-correlated and can propagate across the qubit grid, appearing as dark boxes at coordinates where $x \neq y$. We do not observe these ``bleeding" signatures in the simulation results (Fig.~\ref{fig:memory_expt}, right), which does not treat leakage. 

It is not surprising that leakage is an issue in this experiment. Our circuits did not contain the leakage removal units described in previous work (see e.g. \cite{acharya_suppressing_2023}). As described in the main text, the grafting circuit redefines which sub-lattice of qubits are measured in the middle of the circuit. This is done for compatibility between our color code in the cultivation step and the stabilizers constructed in the circuit in \cite{gidney_magic_2024}. Our readout control parameters for qubit readout were not optimized for this configuration, leading to possible frequency collisions involving qubit $\ket{2}$ states and subsequent leakage. The data-measure re-definition may be avoided in the future, where we start in a different color code stabilizer geometry than reported in the main text, beyond the scope of this work. We expect that improvements in the LER will follow from improved control calibrations when running such a simpler circuit, the addition of leakage removal units, and when extending to larger distances. 
\clearpage

\end{document}